\documentclass[pra,letterpaper, twocolumn]{revtex4-1}
\usepackage[utf8x]{inputenc}
\usepackage{amsmath}
\usepackage{xcolor}%
\usepackage{graphicx}
\usepackage{dcolumn}
\usepackage{bm}
\usepackage{subfigure}
\usepackage{textcomp}

\newcommand{\ket}[1]{\left| #1 \right\rangle}
\newcommand{\bra}[1]{\left\langle #1 \right|}

\newcommand{\abs}[1]{\left| #1 \right|}

\newcommand{\expn}[1]{{\rm e}^{#1}}

\newcommand{\dg}{{^{\dagger}}}

\newcommand{\ie}{\textit{i.e.,}}
\newcommand{\nn}{\nonumber}

\begin{document}
	\title{\textbf{Nanoscale Architecture for Frequency-Resolving Single-Photon Detectors}}
	\author{Steve M. Young$^1$}
	\author{Mohan Sarovar$^1$}
	\author{Fran\c{c}ois L\'{e}onard$^1$*}
	\affiliation{$^1$Sandia National Laboratories, Livermore, CA, 94551, USA}
	\email{fleonar@sandia.gov}

\begin{abstract}
Single photon detectors play a key role across several basic science and technology applications. While progress
has been made in improving performance, single photon detectors that can maintain high performance while also resolving the photon frequency are still lacking. By means of quantum simulations, we show that nanoscale elements cooperatively interacting with the photon field in a photodetector architecture allow to simultaneously achieve high efficiency, low jitter, and high frequency resolution. We discuss how such cooperative interactions are essential to reach this performance regime, analyzing the factors that impact performance and trade-offs between metrics. We illustrate the potential performance for frequency resolution over a 1 eV bandwidth in the visible range, indicating near perfect detection efficiency, jitter of a few hundred femtoseconds, and frequency resolution of tens of meV.  Finally, a potential physical realization of such an architecture is presented based on carbon nanotubes functionalized with quantum dots.
\end{abstract}


\maketitle

\section*{Introduction} 

The efficient detection of single photons is an important capability with wide-ranging uses \cite{Chunnilall:2014}.  Tremendous progress has been made in the development of devices that can attain high efficiency while also achieving minimal dark counts and jitter \cite{Bienfang:2004ij,Woodson:2016cx,Pernice:2012bc,Marsili:2012ib,Marsili:2013th,Eisaman:2011cc, Hadfield:2009}.  Recent work has also focused on imparting new functionality, such as photon-number resolution \cite{Rosenberg:2005, Kardynal:2008, Young:2020}. One functionality that is highly desirable for many applications is the ability to resolve the frequency/wavelength of the detected photon.  For example, a high efficiency single photon detector with frequency resolution would be useful for hyperspectral imaging \cite{Griffiths:2018}, identification of remote galaxies \cite{Aghamousa}, and confocal microscopy\cite{Niwa:2021}.

For detection of light in the classical limit, frequency resolution is straightforward; A typical device operating in this regime comprises a few photon detectors covering different but overlapping energy bands, so that the frequency can be inferred from the relative intensity reported by each detector element.  For example, the human eye uses three types of photoreceptors to span the visible spectrum giving us the ability to distinguish on the order of 10 million colors~\cite{Judd1953}.  However, since this is ultimately a statistical procedure relying on a signal with many photons, this scheme fails in the single photon limit, and an entirely new approach to frequency resolution is required.

The simplest approach for single photon frequency resolution is to spatially guide the photon to different single photon detectors based on its frequency \cite{Kahl:17}, but this becomes more challenging as the number of frequency bins increases. These detectors have achieved frequency resolution $\delta \omega \approx$ 2 meV over bandwidths $\Delta \omega \approx$ 10 meV, timing jitter less than 50 ps, dark count rates less than 10 Hz, at count rates that could attain 100 MHz. For such platforms, the overall detection efficiency is about 19\%.
Further progress has recently been made in this direction by using compact on-chip wavelength dispersion with signal timing information in a meandering superconducting nanowire detector \cite{Cheng:2019} where $\delta \omega \approx$ 2 meV over bandwidths $\Delta \omega \approx$ 3 eV, timing jitter of 41 ps, dark count rates less than 30 Hz, at count rates that could attain 10 MHz. Overall detection efficiency is less than 0.3\%. Other approaches have used tunable electromagnetically induced transmission to perform single photon spectrometry \cite{Ma:17}.

An alternative approach is for one detector to be directly sensitive to the photon frequency. For example, transition edge sensors are sensitive to the total energy of an incoming photon pulse~\cite{Cabrera1998,Rosenberg:2005,Fukuda_2011}, a phenomenon that can be used to extract the photon energy provided there is only one photon in the pulse. Such systems have achieved detection efficiencies greater than 95\%, and could operate at 1 MHz, with an energy resolution $\approx$ 0.2 eV. Recent work \cite{Hattori_2022} has improved the energy resolution to 67 meV, at the cost of a lower detection efficiency of 60\%.

While overcoming a number of engineering challenges could further improve the performance of existing energy-resolving detectors, the above experimental results illustrate the difficulty in achieving high performance across detection metrics. An interesting scientific question is whether it is possible for a detector architecture to simultaneously achieve high performance
in all metrics, and if not, what tradeoffs exist between metrics.

\begin{figure}
	\centering
	\includegraphics[width=0.8\columnwidth]{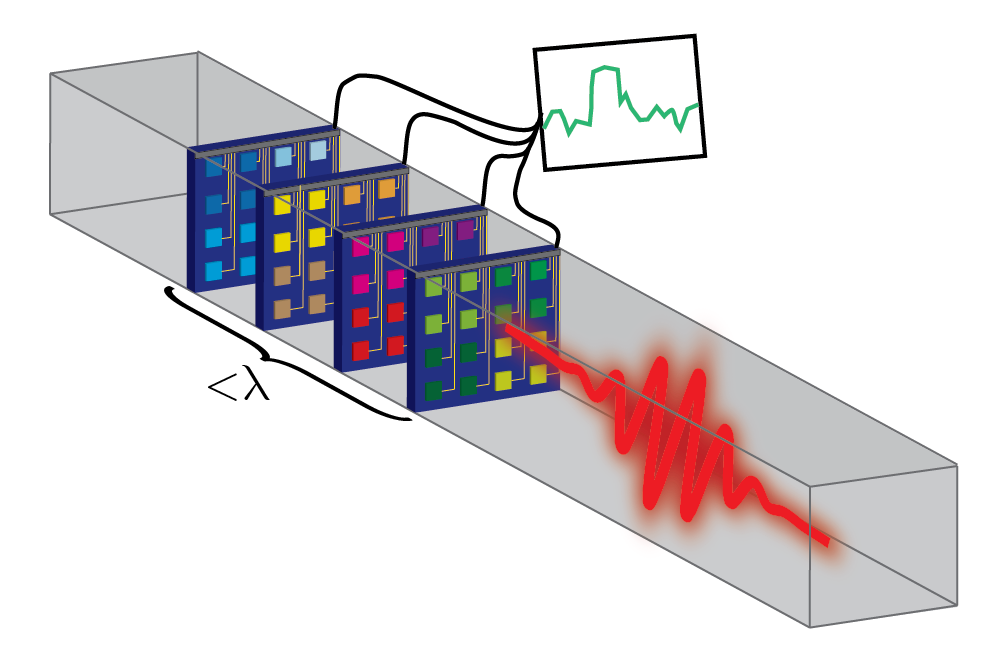}
	\caption{{\bf Illustration of the proposed frequency-resolving single photon detector.} A photon of wavelength $\lambda$ is guided into a single mode waveguide. The subwavelength detector comprises groups of elements, represented by the colored squares, interacting with the photon field and capable of generating a signal when a photon is absorbed (green trace). The elements are coupled not only to the photon but to each other via the field mode, resulting in a collective absorption process.}
\end{figure}

In this work we propose a different approach for a photodetector, schematically shown in Fig.~1, that is capable of accurately determining the frequency of a single photon while also maintaining high efficiency and low jitter. Critical to achieving this performance is the engineered cooperative coherent behavior of detector elements. (We use the term cooperative to refer to the simultaneous interaction of subwavelength detector elements with a common electromagnetic field. Such a situation can lead to superabsorption, but this is not the effect that we will take advantage of in this work.) We discuss the requirements on the general architecture, the resulting performance limits and tradeoffs, and propose a physical realization based on nanoscale materials.

\section*{Results and Discussion}
\subsection*{Design}

Figure 2 shows a schematic of our photodetector design and a corresponding energy level diagram. A single photon of frequency $\omega$ propagates in a waveguide that supports a single mode for frequencies around $\omega_0$. The photon is incident on a detector composed of a collection of sub-wavelength objects arranged hierarchically. The system is divided into $N$ \emph{subsystems} ($N=4$ hexagons in Fig.~2); the $i$th subsystem is made up of $n_i$ \emph{elements} that interact with the photon and an amplifier that produces a signal if one of the subsystem elements absorbs a photon.  Each element (represented by colored squares in Fig. 2) couples to the photon field with strength $\gamma$. Absorption of the photon excites the element from the ground state $0$ to an excited state $1_{im}$, where $m$ indexes the elements in subsystem $i$. The excitation energies $\omega_{im}$ of the elements in subsystem $i$ (which are rendered in the same color in Fig.~2) are centered on a frequency $\omega_i$ associated with the subsystem; the $m$th element's detuning from this frequency will be designated as $\delta_{im}$. We will assume in all cases that the frequencies $\omega_{i}$ are evenly spaced over the desired range of frequency resolution $\Omega$. It is important to note that the entire range $\Omega$ must lie within the single-mode regime of the waveguide; \ie~the width of $\Omega$ must be less than the cutoff frequency of the waveguide. These excited states undergo incoherent decay at rate $\Gamma^2$ to long-lived states $C_{im}$, which once populated remain so indefinitely. These states are monitored and amplified by an output channel (indicated by hexagons in Fig.~2) at rate $\chi$ into a classical signal indicating that the photon was absorbed by system $i$. Thus, the subsystems correspond to frequency bins into which the photon is sorted. 
For detection of single monochromatic photons, the absorption and incoherent rates of a single subsystem can be optimized to yield ideal detection\cite{Young:2018b}. 

Due to the sub-wavelength size of the system, it will exhibit cooperativity\cite{Bienaime2012, Reitz2022} -- the interaction of all the elements with the incident photon is collective (through the field-mediated inter-element interactions).  This type of detector is thus best described as a quantum detector whereby the photon field, the absorption process, and the measurement process are treated as part of one quantum system~\cite{Young:2018b}. This can be done using techniques from quantum optics and quantum information as we now discuss.

\begin{figure}
	\centering
	\includegraphics[width=1.0\columnwidth]{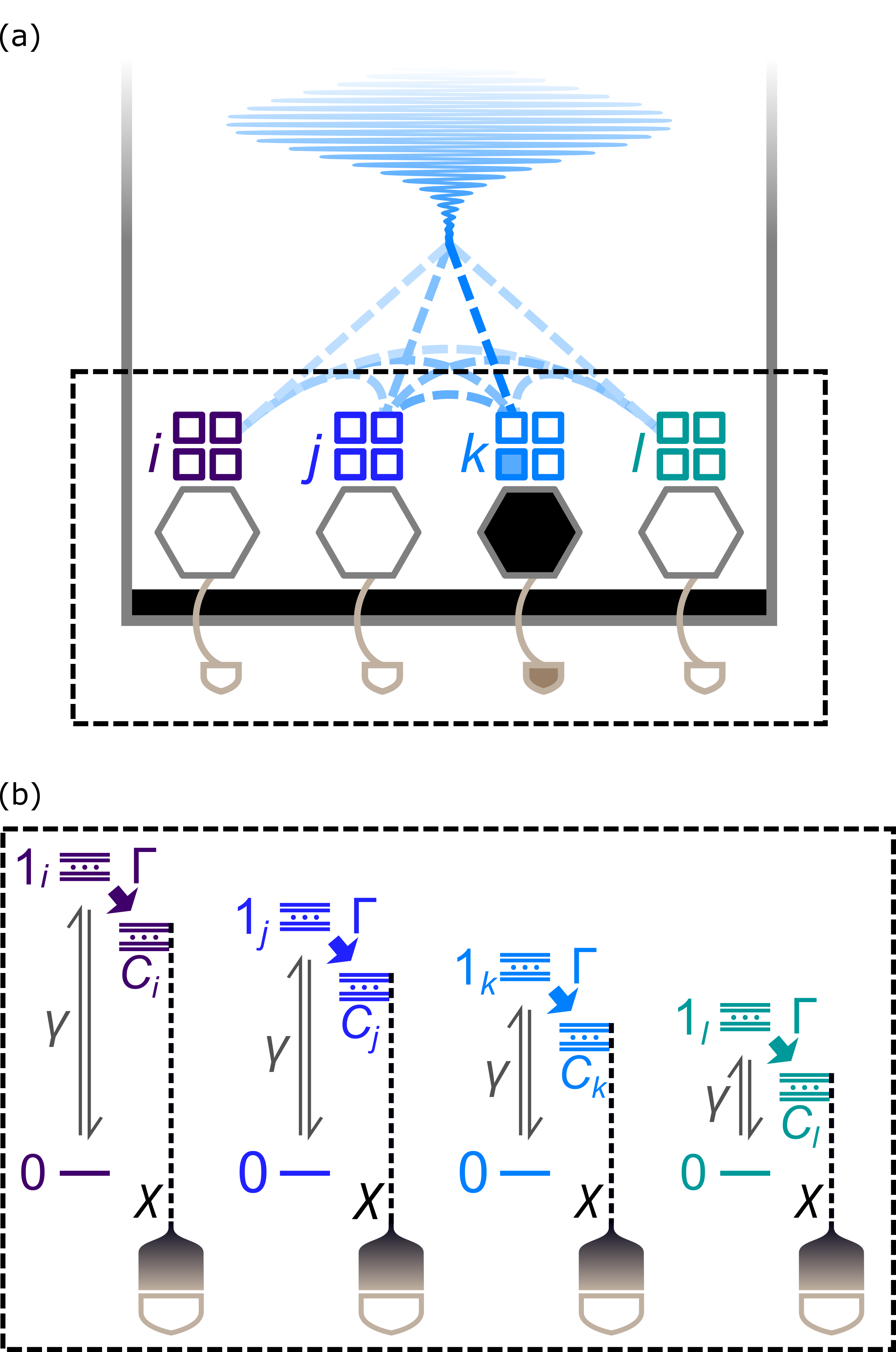}
	\caption{{\bf Internal architecture of the photodetector.}  {\bf a} Illustration of the detector structure for the case of $N=4$. A photon (blue wavepacket) propagating from top to bottom is guided into a single mode waveguide. The detector comprises absorbing elements (squares) and amplifiers (hexagons), which are grouped into subsystems containing multiple absorbing elements that are (near-)degenerate with subsystem-specific transition frequency.  The absorbing elements in all subsystems interact with the photon (as indicated by the dashed lines), which is ultimately absorbed by a single element (filled square), modulating the signal in the associated amplifier (filled hexagon).
		{\bf b} Energy diagram for the detector. The elements are divided into groups characterized by near-degenerate transitions of different energies.  Optical excitation is from the ground state 0 to the excited state 1, followed by incoherent decay to dark states C monitored by amplifiers; each dark state corresponding to elements of different transition frequency are associated with distinct amplifiers allowing for discrimination of the incoming photon frequency.  Dots between lines marking energy states are used to indicate that many more states are present than are drawn explicitly. Here $\Gamma$ is the incoherent transition rate, $\gamma$ is the optical transition rate, and $\chi$ is the measurement rate.}
\end{figure}

	 
\subsection*{Formalism}	
To calculate the properties of the detector we employ a recently developed formalism for modeling quantum photodetection of arbitrary light states~\cite{Young:2018b}. 

The matter-field system that composes the detector is treated as an open quantum system whose density matrix $\hat{\rho}_{\rm TOT}$ evolves according to a quantum master equation.  As shown in \cite{Baragiola:2012cs},
	 for a monochromatic, single-mode wavepacket containing $n$-photons with temporal profile $\varepsilon(t)$ and frequency $\omega_0$, the field degrees of freedom can be eliminated and the matter system density matrix $\hat{\rho}(t)= {\rm Tr}_{\rm LIGHT}\left[\hat{\rho}_{\rm TOT}(t)\right]$ can be evolved according to a hierarchy of equations evolving auxiliary density matrices depending on the initial field state.  We shall confine ourselves to the single photon case with a stable initial state $\hat{\rho}(t_0)$, in which case these can be written using a single auxiliary density matrix $\hat{\varrho}$ as~\cite{Young:2018}
	 \begin{flalign}
	 	\dot{\hat{\rho}}(t)&=\mathcal{V}_{\rm M}(\hat{\rho}(t))+\varepsilon(t)\expn{-i\omega_0 t}[\hat{\varrho}(t),\hat{L}^\dagger]\nonumber\\
	 	&+\varepsilon^*(t)\expn{i\omega t}[\hat{L},\hat{\varrho}\dg(t)]+\mathcal{V}_{\rm L-M, Coop}(\hat{\rho}(t))\nonumber\\
	 	\dot{\hat{\varrho}}(t)&=\mathcal{V}_{\rm M}(\hat{\rho}(t))+\varepsilon^*(t)\expn{i\omega_0 t}[\hat{L},\hat{\rho}(t_0)]\nonumber\\
	 	&+\mathcal{V}_{\rm L-M, Coop}(\hat{\varrho}(t))
	 	\label{eq:full_me}
	 \end{flalign}
 where 
 \begin{flalign}
 	&\mathcal{V}_{\rm M}(\hat{\rho}(t))=-i[\hat{H},\hat{\rho}(t)]+\sum^{\rm BATHS}_{im}\mathcal{D}[\hat{Y}_{im}]\hat{\rho}(t)\nonumber\\
 	&\hspace{3cm}+\sum^{\rm AMPS}_i\mathcal{D}[(2k_i)^{1/2}\hat{X}_i]\hat{\rho}(t)\nonumber\\
 	&\mathcal{V}_{\rm L-M, Coop}(\hat{\rho}(t))=\mathcal{D}[\hat{L}]\hat{\rho}(t).
 \end{flalign}
Here $\mathcal{D}$ represents the Lindblad superoperator $\mathcal{D}[\hat{O}]\hat{\rho}=\hat{O}\hat{\rho}\hat{O}\dg-\frac{1}{2}\left\{\hat{O}\dg\hat{O},\hat{\rho}\right\}$ and we have assumed $\hbar=1$. The matter system is thus governed by an internal Hamiltonian $\hat{H}$, its coupling to the optical field captured by the operator $\hat{L}$, its coupling to external reservoir(s) (BATHS) captured by operator(s) $\hat{Y}_{im}$, and the dynamics induced by the measurement/output channels (AMPS) which couple into the system with rate(s) $2k_i$ and operator(s) $\hat{X}_i$. We note that the cooperative interactions are captured by the term $\mathcal{V}_{\rm L-M, Coop}$ and are furnished by the $\hat{L}$ operator via the Lindbladian term; $\hat{L}\dg\hat{L}$ contains, in addition to diagonal matrix elements corresponding to single system element spontaneous emission, off-diagonal matrix elements that couple system elements.  As shown previously\cite{Young:2018b, Mattiotti_2022}, cooperative interactions can play a crucial role in single photon detection and must be accounted for in optimizing detector parameters.  In particular, the distributed absorption over detector elements allows for longer overall collection and measurement processes and sharply defined detection bandwidths that are crucial to high performance frequency resolution. For our design these operators are:
\begin{flalign}
	\hat{H}&=\sum_{i}^N\sum_{m}^{n_i}\left(\omega_{i}+\delta_{im}\right)\ket{1_{im}}\bra{1_{im}}\nn\\
	\hat{L}&=\sum_{i}^N\sum_{m}^{n_i}\gamma\ket{0}\bra{1_{im}}\nn\\
	\hat{Y}_{im}&=\Gamma\ket{C_{im}}\bra{1_{im}}\nn\\
	\hat{X}_i&=\sum_{m}^{n_i}\chi\ket{C_{im}}\bra{C_{im}}.\label{eq:model}
\end{flalign}
The $\omega$ quantities are assumed to have units of energy, while $\gamma$, $\Gamma$, and $\chi$ have units of square root energy and are associated with rates. The initial state of the detector, $\hat{\rho}_{\rm M}(t_0)$, is assumed to be the ground state of all absorbing elements and the photon wavepacket is assumed to contain a single photon ($n=1$).  We note that practically there will be element-wise variations in parameters as well as additional processes present due to impurities, disorder, and other non-idealities.  However, for the sake of numerical tractability and clarity we restrict ourselves to this simplified model. We also neglect spatial variations of the photon mode in the waveguide; the optimization approach described below could include this detail. In this model we omit the decay from the $C$ state to the ground state as it was previously shown that such a system can be engineered for high performance monochromatic single photon detection.

In general one can write measurement outcomes $\Pi(t)$ as
\begin{flalign}
&\Pi(t) = {\rm Tr}\left[\mathcal{K}(t,t_0)\hat{\rho}_{\rm TOT}(t_0)\right],\label{eq:basic2}
\end{flalign}
where $\mathcal{K}$ is an operator that accounts for both the detector state evolution and measurement output collection and processing, which can be used to determine detector performance\cite{Young:2018b}. In cases such as the present one, where the $C$ states are stable (there is no population transfer to other states), this takes a simple form and we can write the probability that a single incident photon of frequency $\omega_0$ registers as a hit in output channel $i$ at time $t$ as \cite{Young:2018b,Young:2020}
\begin{flalign}
	\Pi_{i}(t;\varepsilon_{\omega_0}) =& {\rm Tr}\left[\hat{x}_{i}\hat{\rho}(t;\varepsilon_{\omega_0})\right]
	\label{eq:hitprob}
\end{flalign}
where $\hat{x}_i$ is a projection matrix such that $\chi\hat{x}_i=\hat{X}_i$.

From this formalism we can determine a number of key metrics, of which the following are presently of interest:

{\it Efficiency}. Efficiency is the probability of having detected the photon after it has passed the detector; \ie~ at $t\rightarrow \infty$.  For a given channel this probability is directly given by $\Pi_{i}(\infty;\varepsilon_{\omega_0})$, for which we use the simplified notation $\Pi_{i}(\varepsilon_{\omega_0})$ while for the overall efficiency we write $P(\varepsilon_{\omega_0})= \sum_{i}{\Pi_{i}(\varepsilon_{\omega_0})}$. In the present work we assume $\varepsilon(t)$ to be infinitely broad, corresponding to a Delta function in frequency space at $\omega_0$. We have previously shown that this is a good approximation even for pulses that are short in a practical sense \cite{Young:2018}.  In this case these quantities become functions of the frequency $\omega_0$ only.  Furthermore, since $\hat{Y}_{im}\hat{Y}_{jk}=\hat{Y}_{im}\hat{L}=\hat{Y}_{im}\hat{X}_i=0\,\forall i,j,k,m$\cite{Young:2020}, we obtain from Eq.~\eqref{eq:hitprob}:
\begin{flalign}
	\Pi_{i}(\omega_0)=&\sum_{m=1}^{n_i}\abs{\bra{C_{im}}\hat{Y}_{im}\left[i(\omega_0-\hat{H}-\hat{H}_D)\right]^{-1} \hat{L}\dg\ket{0}}^2\nn\\
	\hat{H}_D=&i\frac{1}{2}\left(\sum^{\rm BATHS}_{im}\hat{Y}_{im}\dg\hat{Y}_{im}+\hat{L}\dg\hat{L}\right)
\end{flalign}
as the long-time or steady-state probability that output channel $i$ registers the photon.   

{\it Jitter}. Since $\dot{\Pi}(t;\varepsilon_{\omega_0})$ gives a distribution of detection times, the jitter for channel $i$ is
\begin{flalign}
	\sigma_i(\varepsilon_{\omega_0})&=\sqrt{\int_{t_0}^\infty dt\, t^2\frac{\dot{\Pi}_i(t;\varepsilon_{\omega_0})}{\Pi_{i}(\varepsilon_{\omega_0})}-\left(\int_{t_0}^\infty dt\, t\frac{\dot{\Pi}_i(t;\varepsilon_{\omega_0})}{\Pi_i(\varepsilon_{\omega_0})}\right)^2}
\end{flalign}
with the total jitter obtainable by replacing $\Pi_i$ in the above with $P$.
Since the overall jitter strongly depends on the pulse characteristics, we define $\sigma_{{\rm SYS}}$ as the total jitter minus the temporal width of the pulse

\begin{flalign}
	\sigma_{{\rm SYS}}(\varepsilon_{\omega_0})&=\sqrt{(\sigma(\varepsilon_{\omega_0}))^2-\left(\sigma_0\right)^2}
\end{flalign}

with
\begin{flalign}
	\sigma_0&=\sqrt{\int_{t_0}^\infty dt\, t^2\abs{\varepsilon(t)}^2-\left(\int_{t_0}^\infty dt\, t\abs{\varepsilon(t)}^2\right)^2}.
\end{flalign}
The regime of infinitely broad $\varepsilon(t)$ under consideration corresponds to the limit $\sigma_0\rightarrow \infty$, in which $\sigma(\varepsilon_{\omega_0})$ goes to infinity as well.  However, $\sigma_{{\rm SYS}}(\varepsilon_{\omega_0})$ remains finite and converges to a fixed value which can be obtained numerically; we will thus report the jitter defined as
\begin{flalign}
	\sigma_{{\rm SYS}}(\omega_0)&=\lim_{\sigma_0\rightarrow \infty}\sqrt{\left(\sigma(\varepsilon_{\omega_0})\right)^2-\left(\sigma_0\right)^2}
\end{flalign}

{\it Frequency Resolution}.
The $\Pi_i$ furnish a set of probabilities that an incident photon of frequency $\omega_0$ will be recorded as photons of frequency $\omega_i$.  We can thus write the expected measurement frequency and standard deviation as
\begin{flalign}
	\omega_\mu(\omega_0)&=\sum_{i}\omega_i\Pi_i(\omega_0)/P(\omega_0)\nn\\
	\omega_\varsigma(\omega_0)&=\sum_{i}(\omega_i-\omega_\mu)^2\Pi_i(\omega_0)/P(\omega_0).
\end{flalign} 
We will use the latter to define frequency resolution.

\begin{figure*}
	\centering
	\includegraphics[width=2.0\columnwidth]{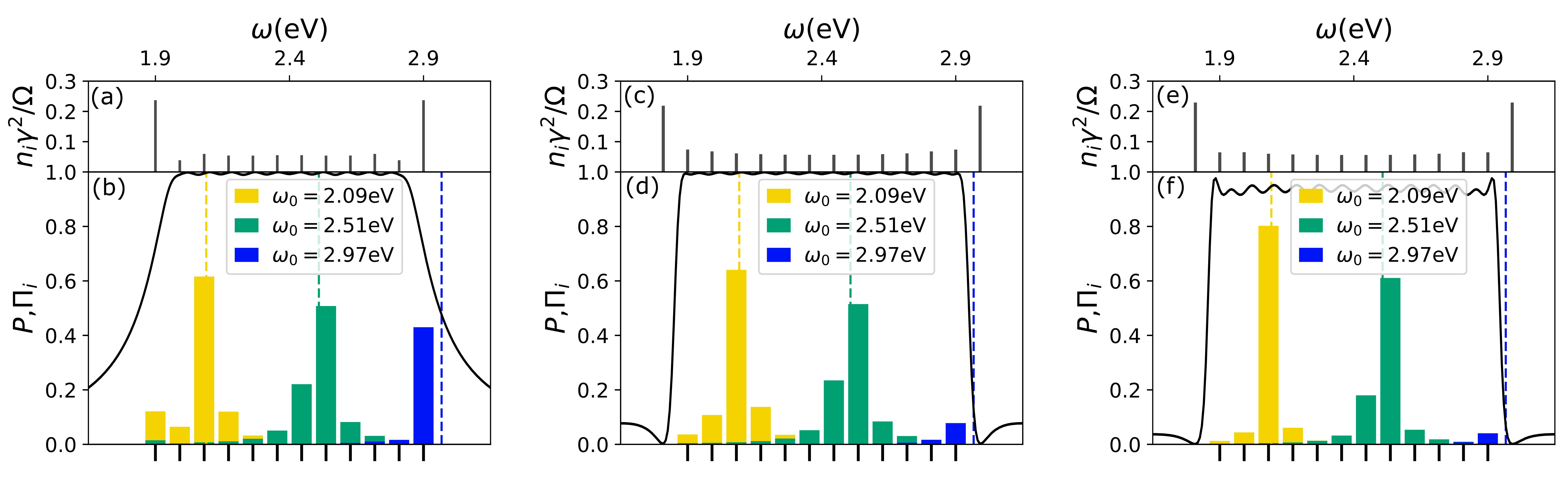}
	\caption{{\bf Performance of photodetector.} {\bf a,b} Implementation for $N=12$ frequency bins when all the absorbers are monitored. Here the incoherent transition rate $\Gamma^2 = 0.085eV$. {\bf c,d}. $N=12$ system where end subsystems that do not decay to monitored dark states are added. Here the incoherent transition rate $\Gamma^2 = 0.082eV$. {\bf e,f} A version of the design of panels  ({\bf c,d}) where the incoherent decay rate $\Gamma^2$ is reduced by half, resulting in higher frequency resolution but reduced overall efficiency. For each configuration, photons of three frequencies $\omega_0$ are considered, denoted by different colors and marked by dashed lines. In panels ({\bf b,d,f}) the colored bars indicate the probability of a photon of that frequency $\omega_0$ being detected at each bin, the characteristic frequencies of which are marked by the x-axis ticks.
		\label{fig:eff}
	}
\end{figure*}

\subsection*{Theoretical Performance}
Given the above design and model, we now optimize the parameters in the model (Eq.~\eqref{eq:model}) to achieve optimal tradeoff between efficiency, frequency resolution and jitter. Both the parameters and device metrics can be defined in terms of the width of $\Omega$, which sets a frequency scale for the model.  For simplicity we will set $\Omega$ to be the range $(1.9,2.9)$ eV, covering the bulk of the visible spectrum. 
For a given set of parameters $N,\gamma$, and $\Gamma$, we take $n_i$ -- the number of elements in subsystem $i$ -- to be optimization parameters. These must be optimized for performance due to their collective interaction with the field. While for some systems analytical expressions for this optimization are available \cite{Young:2018b}, in the present case it must be done numerically. In what follows we will use the set of $n_i$ such that $\max_{-\Omega/2<\omega_0<\Omega/2} [1-P(\omega_0)]$ is minimized, i.e.\ the worst case inefficiency over the detector bandwidth is minimized. This minimization was performed using the L-BFGS-B algorithm~\cite{Byrd95}. We note that under the described conditions the results are independent of the parameter $\chi$ as discussed in previous work \cite{Young:2018}.

The results are shown in Fig.~3(a,b) for $N=12$, with the $\omega_i$ equally spaced from 1.9 eV to 2.9 eV (the results can be extended to larger $N$ but become computationally more demanding. However we note that the efficiency and jitter are essentially independent of $N$, so that the main role of $N$ is to impact the frequency resolution). Figure 3(a) shows that the optimized $n_i$ (scaled by $\gamma^2/\Omega$ to remove the dependence on parameters $\gamma$ and $\Omega$) for each subsystem follow a non-uniform distribution across the detection range. This distribution with peaks at the end of the range is reminescent of the density of states in quasi-one-dimensional systems which was previously shown to ensure optimal efficiency over a broad frequency range in non-frequency-resolving detectors \cite{Young:2020}. As shown in Fig.~3(b) the efficiency of detecting the photon is at least 99\% over the target range, with relatively narrow distributions; the $\omega_\varsigma$ for the three sample frequencies are 133 meV, 145 meV, and 112 meV.   

One non-ideal aspect of Fig.~3(a) is the excess absorption at the ends, especially for frequencies outside the desired spectral range. For example, the blue photon with frequency outside the detection range still gives an apparent peak in one of the detection channels, with a 57\% probability of being detected at $\omega_\mu=2.872$ eV.  This arises from the large $n_i$ necessary at the ends to obtain uniformly high efficiency~\cite{Young:2020}. To combat this effect we include elements uncoupled to decay channels and amplifiers at 1.81 eV and 2.99 eV. This modulates the field coupling of the overall detector while substantially reducing the absorption outside the desired band (Fig.~3(c,d)) and removes the need for large $n_i$ in highest and lowest frequency bins. As a result, the purple photon for example no longer appears with a high probability in detection bins, being detected only 10\% of the time.  This effect highlights the cooperative nature of the detection; absorptive elements outside the frequency range of desired detection can be exploited to shape the detection frequency window.

The system can also be engineered to improve the frequency resolution. In Fig.~3(e,f) we show a system with an incoherent decay rate $\Gamma^2$ reduced by half which leads to a narrower distribution of detection probabilities around the target bin. This arises because the slower decay rate leads to a reduced broadening of the absorption spectrum. However, this reduced broadening prevents full coverage of the detection window, leading to a reduced overall detection efficiency. Larger $N$ can reduce the frequency spacing of subsystems, allowing more complete coverage of the frequency range $\Omega$, but at the cost of increased device complexity.  Thus, for this particular architecture there appears to be a trade-off between efficiency and frequency resolution for a given $N$.

\begin{figure}
	\centering
	\includegraphics[width=1.0\columnwidth]{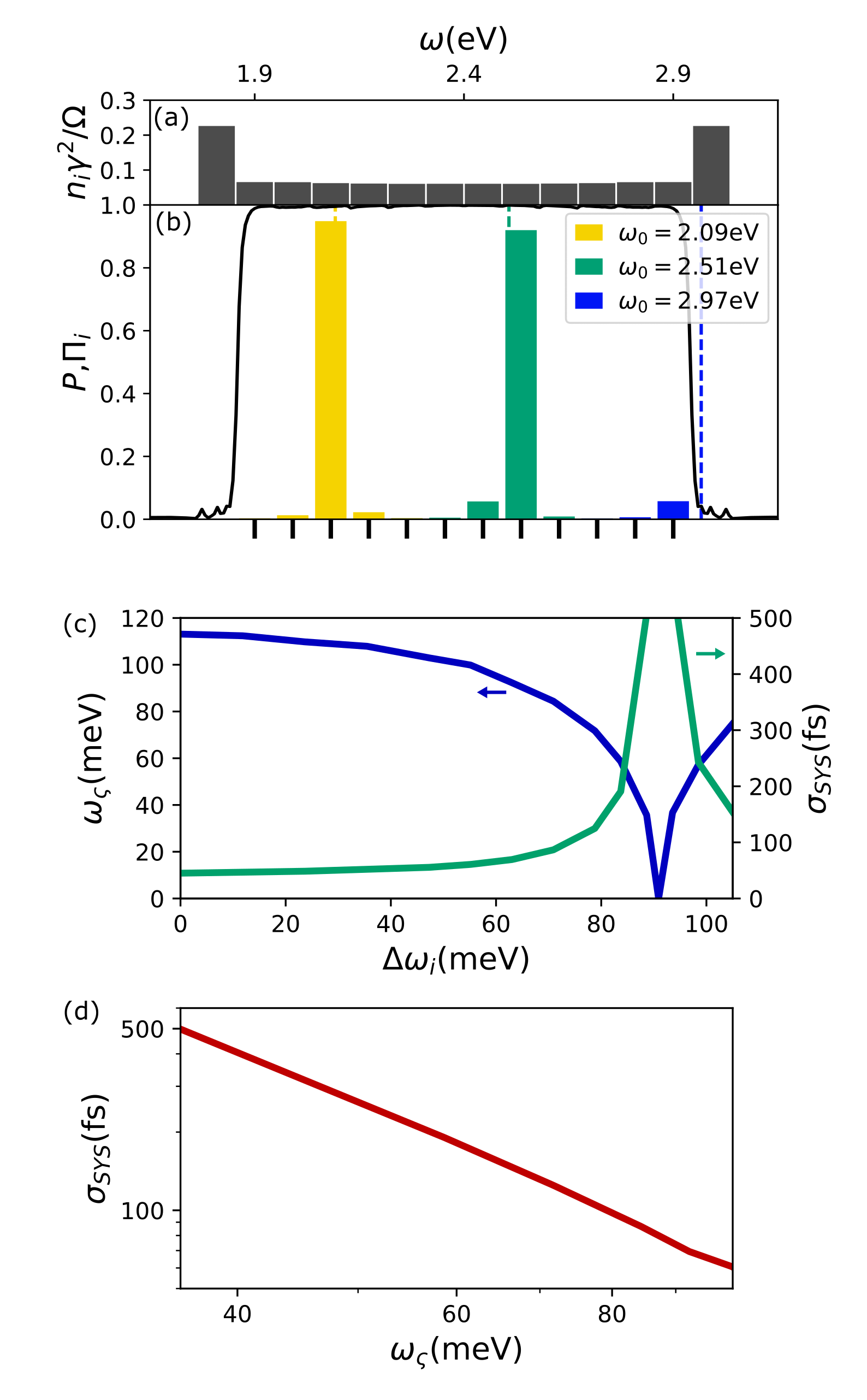}
	\caption{{\bf Simultaneous high performance across metrics.} {\bf a,b} Performance of the system with the absorption of each subsystem elements uniformly spread over a range of 88.6meV, reflected in the width of the grey bars in ({\bf a}). 
In ({\bf b}) the response for photons of three differencies $\omega_0$ is shown, indicating very narrow frequency resolution while maintaining high efficiency.  {\bf c} $\sigma_{{\rm SYS}}$ and $\omega_\varsigma$ that result from the slowest $\Gamma^2$ that maintains greater than 99\% efficiency for a given $\Delta\omega_i$. {\bf d} $\sigma_{{\rm SYS}}$ plotted against  $\omega_\varsigma$ from panel ({\bf b}) on a log/log scale.
		\label{fig:dispersion}}
\end{figure}

An alternative approach to improve frequency resolution is to
introduce dispersion in the transition energies of subsystem elements.  We will take this dispersion to be flat over a range $\Delta\omega$, such that $\delta_{im}\in (-\Delta\omega/2,\Delta\omega/2)$.  In Fig.~4(a-d) we show results for the case where the subsystem elements were given transition energies spread over $\Delta\omega=88.6$meV  -- slightly less than the bin frequency spacing -- and $\Gamma$ was optimized to ensure $\ge 99$\% efficiency over the specified frequency range.

The result is improved frequency resolution as evidenced by the higher central peaks in each bin. The impact of the absorption spread on frequency resolution is plotted in Fig. 4(c), showing how it is minimized with increasing $\Delta\omega$ until it reaches the bin frequency spacing. This result was obtained by choosing $\Gamma$ for each $\Delta\omega$ to be as slow as possible while achieving efficiency $\ge 99$\%.

Thus, the tradeoff between efficiency and resolution has been essentially eliminated. In addition this design also gives low jitter. Indeed, Fig.~4(c) shows the relationship between frequency resolution and jitter as a function of the bin width. For small bin widths, the jitter is as low as 50 fs with a frequency resolution of 115 meV. As the bin width increases, the jitter remains low as the frequency resolution improves and starts to increase as further improvements in frequency resolution are obtained for larger bin widths. This constitutes a trade-off between jitter and frequency resolution, as shown in Fig. 4(d). In practice this is not particularly onerous since a 35 meV frequency resolution still only has 500 fs of jitter. The tradeoff arises because on the one hand $\Gamma$ determines the jitter (with higher values of $\Gamma$ giving lower jitter), while small values of $\Gamma$ are required to reduce broadening and maximize the frequency resolution. We also note that a non-uniform distribution of states over the bin energy (e.g.~gaussian) will generally worsen the frequency resolution, as it is akin to improperly assigning elements near the edges of one bin to the adjacent one. Additionally, depending on the distribution, it may increase the corrugations in the overal efficiency. The later could be compensated for by increasing the incoherent rate $\Gamma$, though at the expense of some additional jitter.
\begin{figure}
	\centering
	\includegraphics[width=1.0\columnwidth]{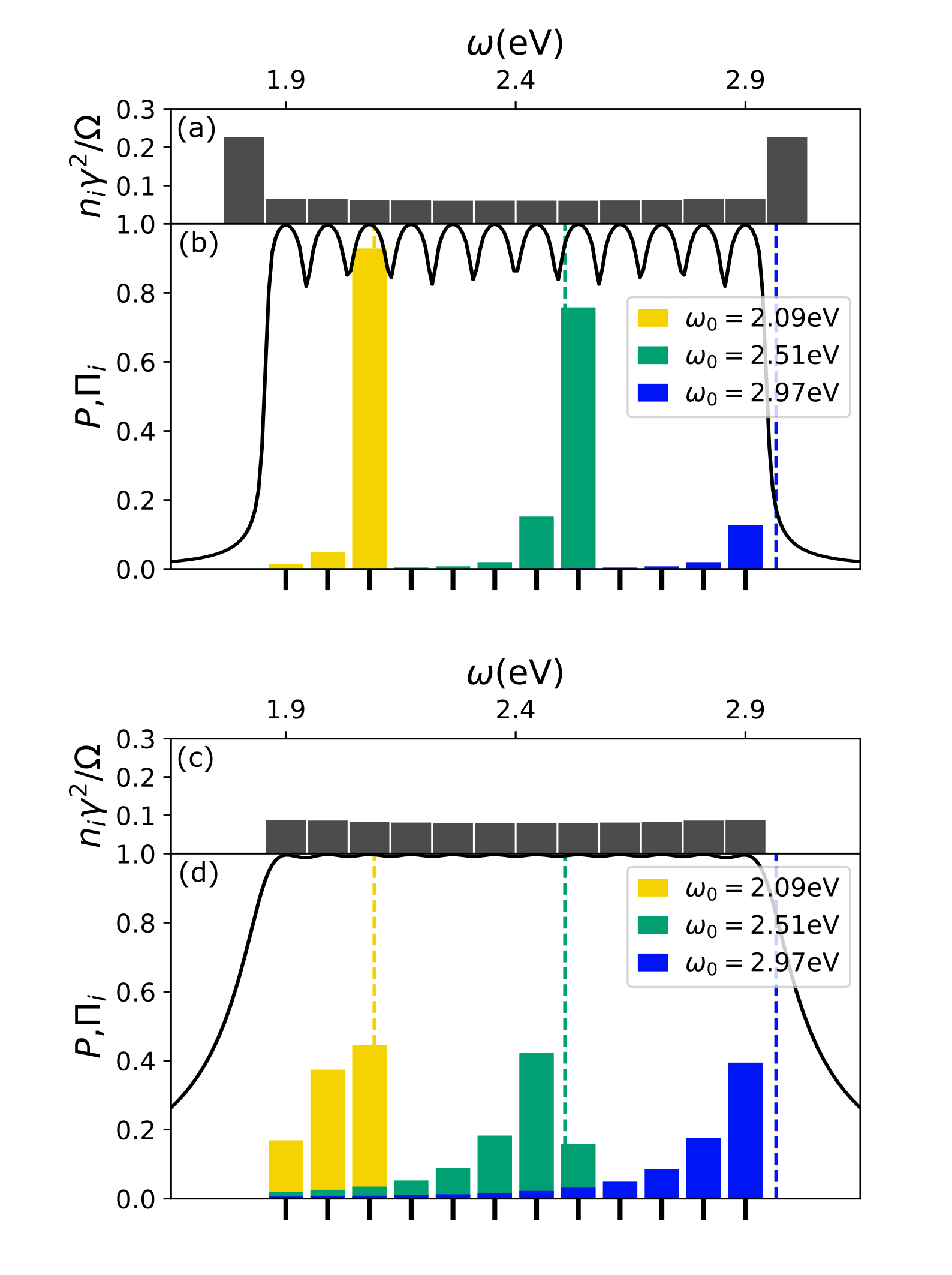}
	\caption{{\bf Performance of photodetectors without cooperativity.} Performance of detectors comprising independent subsystems interacting with the photon in sequence, starting from the lowest frequency bin.  Under these conditions the nonabsorbing lowest and highest frequency subsystems have no impact on the detector performance and are omitted. {\bf a,b} The subsystem compositions are the same as the optimal detector when cooperativity is included.  {\bf c,d} The subsystems are calibrated to satisfy the condition of efficiency $\ge 99$\% over the detector frequency range.  In panels ({\bf b,d}) the colored bars indicate the probability of a photon of that frequency $\omega_0$ being detected at each bin, the characteristic frequencies of which are marked by the x-axis ticks.
		\label{fig:independent}}
\end{figure}

In order to highlight the importance of cooperativity to detector performance, first consider the case of fully independent detectors. Previously\cite{Young:2018}, we showed that an engineered two-level system could function as a perfect narrowband single photon detector, and therefore it should be possible to realize a high performance frequency-resolving detector by sequentially organizing such detectors in a waveguide and separating them by more than a wavelength. The challenge in this case is the large number of detectors needed to cover the detection bandwidth of interest; indeed, as discussed below, the absorbtion width for two-level systems is on the order of $\mu$eV  so a large number of detectors would be needed to achieve uniform coverage over a bandwidth of interest. For example, a 1 eV bandwidth would require one million detectors which would occupy at least 50 cm for light in the visible range. Cooperative effects allow us to engineer the light-matter interaction in the detector in order to circumvent these limitations. This is possible because even non-resonant elements influence the interaction of the resonant elements with the field.

To further illustrate the role of cooperativity, we also performed simulations for the case where the absorbing elements for the different frequency bins are confined to different planes, with the frequency planes separated by more than the photon wavelength (Fig.~5(a-d)). Thus, the system consists of independently absorbing planes, within which cooperative effects exist.  The subsystems are taken to interact with the photon in sequence, starting from the lowest frequency bin; the photon interacts with a given subsystem only if it is not detected by the prior subsystems.  If the subsystems from Fig.~4 are used then the frequency resolution is nearly as good, but the overall efficiency away from the center frequency of the bins suffers, falling below 90\% at the midpoints between bins (Fig.~5(b)).  On the other hand, if $\Gamma^2$ is increased to satisfy the $\ge 99$\% efficiency condition over the whole frequency range, then frequency resolution is significantly compromised (Fig.~5(d)).  In addition, the absorption strength $n_i\gamma^2$ of each bin must be higher.  It might be possible to re-engineer the density of states within each independent bin to re-establish high performance, but this would essentially rely on cooperative effects. Thus, cooperativity provides clear advantage over independently interacting systems.

Ultimately, this analysis reveals that an appropriately designed and constructed detector can achieve high efficiency, low jitter, and arbitrarily fine frequency resolution, with tradeoffs appearing only at performance extrema.  The main challenge that remains is to choose materials and methods allowing for sufficiently precise fabrication for the desired performance regime. 

\subsection*{Physical realization}

In this section, we discuss how the optimal design of Fig.~4 can be physically realized. We consider components confined in a single-mode waveguide like the one in Fig.~1 with detector elements arranged in multiple layers (Fig.~6(a-d)). For the basic components, we focus on carbon nanotubes (CNTs) functionalized with quantum dots (QDs), since this approach has been experimentally shown to give ultrahigh responsivity at room temperature \cite{bergemann, Weng_2015} for classical light fields. In addition, approaches have been demonstrated for controlling the density of QDs around the CNTs \cite{Ka_2012} and for integrating CNTs functionalized with different QDs in the same electronic platform \cite{Ye_2021}. Furthermore, detailed non-equilibrium quantum transport simulations have been employed for in-depth simulations of functionalized CNT devices for detection of monochromatic single photons \cite{Spataru_2019, Leonard_2019, bergemann}, and their connection to the formalism employed here has been presented \cite{Young:2018}.

In the implementation considered here, the photon is absorbed by a QD, with the QD exciton state serving as the excited state of the two-level system in our model. We assume that the QD shape is nearly spherical so that the sensitivity to the photon polarization is minimal, or that the QDs are oriented to maximize the absorption for the propagating mode in the waveguide. The incoherent decay pathway is furnished by exciton dissociation, with either the free electron or free hole being transferred to the CNT and conducted away, while the remaining charge modulates the electronic transport in the CNT. Frequency resolution is enabled by having multiple CNT devices in the waveguide, each functionalized with QDs of different exciton absorption energies. The CNT devices are stacked in the waveguide to improve absorption and frequency coverage. Above and below these planes, unmonitored layers of QDs are added to control the absorption outside of the range of interest. We note that in this case, depicted in Fig.~6(c), it is necessary for QDs to be adjacent to CNTs, and for QDs adjacent to different CNTs to be electronically uncoupled, preventing carrier transfer to other QDs. In the case of the unmonitored layers, all QDs must be uncoupled. The degree of coupling -- or lack thereof -- can be controlled by tuning the distance and level of contact between the quantum dots~\cite{Cui2019}.

The use of QD exciton states constrains the range $\Omega$ over which detection can occur: in order to behave properly as two level systems, the exciton absorption energy corresponding to the lowest energy bin must be separated from the continuum absorption edge of the dot by more than $\Omega$ at least.  This difference is constrained by the binding energy of the exciton, which may in some cases be as high as 1eV, but can be lower depending on the QD size.  We will consider $\Omega=0.4$ eV in the following for frequencies between 2.2 eV and 2.6 eV which is typical of QD systems such as CdSe \cite{Meulenberg:2009}. In this case the QD diameter varies between 3-5 nm.

\begin{figure*}
	\centering
	\includegraphics[width=1.5\columnwidth]{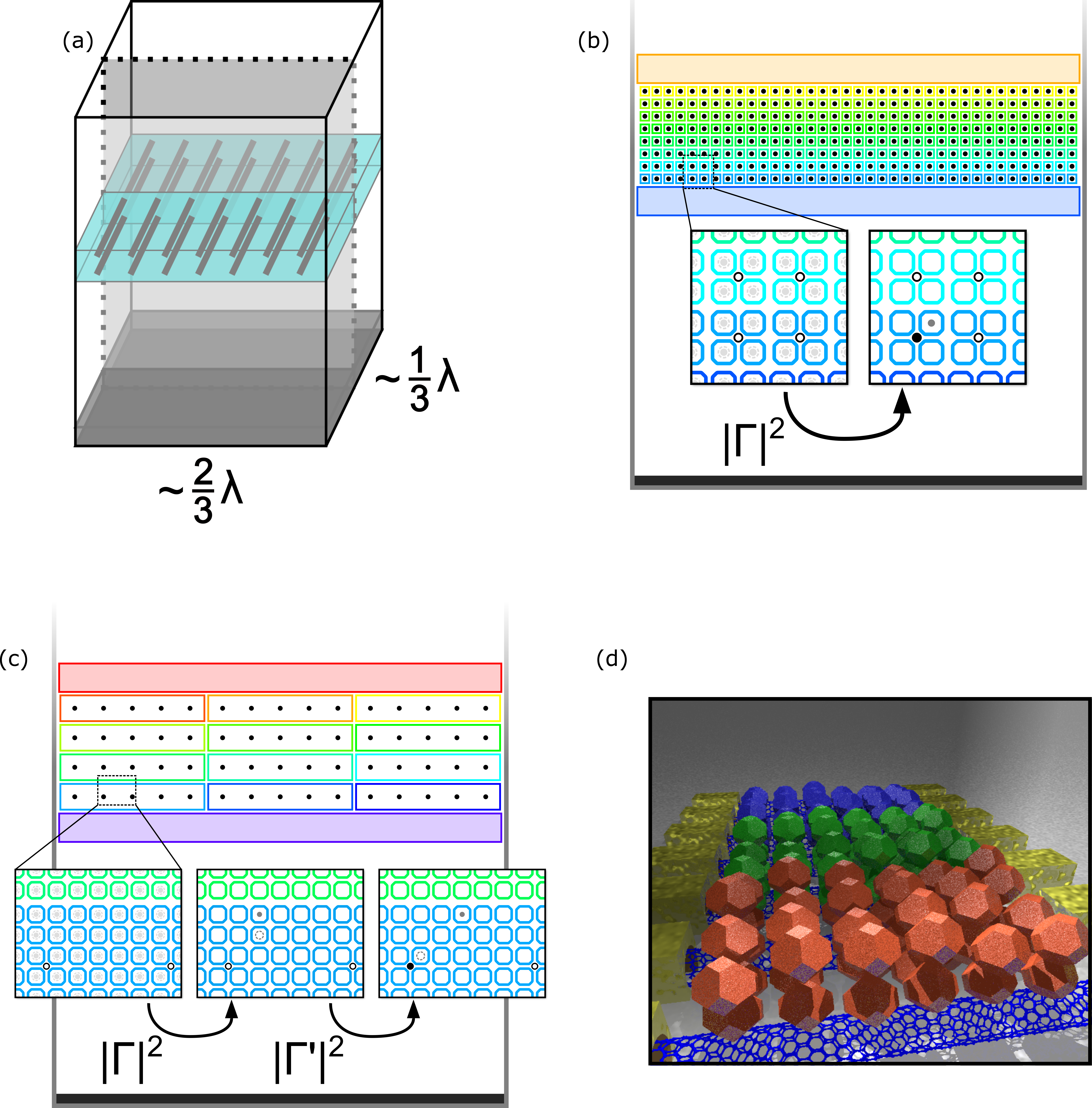}
	\caption{{\bf Physical realization of photodetector.} {\bf a} The device as situated in the waveguide.  The teal region is the active area of the device containing absorbing and measuring components.  The dark grey region represents a mirror at the end of the waveguide. The vertical plane with a dotted outline shows the cross section taken for depiction of the two device configurations shown in panels
		({\bf b}) and ({\bf c}).  In these the quantum dots (QDs) are shown as colored squares, with the carbon nanotube (CNT) measurement channel cross section shown as a black circle. $\lambda$ is the wavelength of the incoming photon. {\bf b} The most direct realization of the design.  The incoherent process in this case (depicted in the inset with rate $\Gamma$) is the separation of the QD exciton, depicted as a solid gray circle for the electron and a dotted line circle for the hole, with a carrier (in this case the hole) migrating into the CNT.  The field due the remaining carrier alters the conductivity of the CNT (indicated by the black filling), which is detected as an absorption event.  {\bf c} A transduction process comprising multiple steps.  The first part of the incoherent decay is the separation of the exciton (rate $\Gamma$), followed by migration to adjacent quantum dots (rate $\Gamma'$).  A field near the carbon nanotube drives the carriers apart, with one arriving at a dot adjacent to the tube, where it modulates the CNT conductivity. {\bf d} A more detailed representation of the CNTs with the QD functionalization is shown. \label{fig:realization}}
\end{figure*}

A waveguide cross-section for this frequency band is about 400 nm$\times$200 nm in size; assuming that the CNT spans the whole waveguide with the electrodes outside of the waveguide, the full 200 nm of the CNT length is available for functionalization.  For the average QD diameter of 4nm, each CNT would have 200 QDs if the QDs are densily packed around the CNT. Each CNT device would occupy about 8nm$\times$8nm in cross section, implying that 50 devices could fit across the length of the waveguide, and that a layer of functionalized CNTs would be about 9-10nm thick including spacing for isolation and contain about 10000 QDs.  If each layer is assigned to a specific wavelength, then all the CNTs in the layer can be connected to the same source/drain electrodes, which can be 10 nm in thickness with 15 nm pitch. The condition that the stack thickness fall well within a wavelength ($<1/3\lambda$) implies that $N=8$ is roughly the number of frequency bins that the detector could support, assuming that two unmonitored layers are added above and below. 

In this case, each subsystem requires a narrow absorption peak of width $\omega_i=\Omega/8=50$ meV.  
To determine if this arrangement can achieve high performance, we need to estimate the quantity $n_i \gamma^2/\Omega$ and compare with the values $n_i\approx0.05$ in the top panel of Fig.~4a. The value of $\gamma$ for QDs in the waveguide is on the order of the free-space spontaneous emission \cite{Young:2020}; QD radiative lifetimes have been measured to be as short as 200ps~\cite{Raino2016}, giving $\gamma^2=3.2 \mu eV$. Thus, for the above design we obtain $n_i \gamma^2/\Omega$ = 0.08, suggesting that the basic design could attain high performance.

One limitation of the basic design is the need for each QD to be in contact with a CNT.  In addition to limiting the number of QDs due to inefficient packing, this also demands many CNT channels and precise fabrication.  This may be ameloriated by introducing additional QDs that are not in direct contact with the CNT, but that can transfer their excitation to another QD adjacent to the CNT. In this scenario, shown in Fig.~6(c), the carrier is blocked from migrating into the CNT channel, and multiple shell of QDs may be associated with a single nanotube; a layer of the system will take the form of layers of QDs with CNTs running through the center. A layer about five dots thick ($\sim 20$nm) would then contain around $3n_i$ dots (assuming $\sim20$\% tighter packing than the first case); partitioning this layer into three subsystems would require four layers, or 90 nm including spacing between subsystem layers.  In this case we could take $\Omega=0.6$ eV and $N=12$, with the unmonitored bins above and below requiring around 50 nm, and remain under our depth budget. Thus the more flexible transduction process allows greater QD density -- and ultimately greater frequency range -- as well as simpler fabrication.  This comes, however, at the cost of increased jitter due to a varying number of additional steps associated with the carrier migration.

Fabrication of the proposed design is challenging, but several fundamental demonstrations make it plausible. As mentioned above, photodetectors with QD-functionalized CNTs have been demonstrated. In addition, waveguide-integrated CNT photodetectors have been realized \cite{Riaz_2019} including with dense arrays of CNTs in the waveguide \cite{Ma_2020}. In terms of addressing individual devices at high density, nanometer size low-resistance contacts to CNTs have been demonstrated \cite{Cao_2017}, while e-beam lithography has been extended to 10 nm pitch \cite{Manfrinato_2017}. Other approaches could also be employed to control the absorption frequency of each element, such as putting molecules or atoms in electric field gradients.

\section*{Conclusion}

We propose a design for a single photon detector capable of intrinsically resolving frequency while maintaining high efficiency and low jitter.  The challenge of doing so is distinctly greater than for generic light, since averages over many photons are not available. Our theoretical analysis clarifies the technical challenges that must be overcome in order to realize such a device, as well as fundamental limitations, and highlights the critical role of cooperativity in achieving optimal performance. As a specific example, we find that frequency resolution of tens of meV over a 1 eV bandwidth is possible while achieving near perfect detection efficiency and jitter of hundreds of femtoseconds. The required design is shown to require a non-trivial distribution of absorbing elements in each frequency bin, reminescent of the density of states in quasi-one-dimensional systems. Our design dictates the need for precision nanoscale engineering capabilities in order to exploit cooperativity and ensure consistent and reliable frequency discrimination in addition to efficient detection. While the precision needed to realize our detector design is demanding, it is not out of reach of modern nanoscale engineering technologies. Moreover, our design represents a benchmark to aim for, and evidence that simultaneous optimization of efficiency, jitter and frequency resolution is possible in photodetection. It also demonstrates the utility of quantum optics and quantum information formalisms to understand the ultimate limits of photodetection, and opens up a path for studying even more complex detectors, such as those that could simultaneously perform photon number resolution and frequency resolution.

\section*{Data availability}

The data that support the findings of this study are available from the corresponding author upon reasonable request.

\section*{Code availability}

The code used for obtaining the presented numerical results is available from the corresponding author upon reasonable request.

\section*{Acknowledgments}

We thank the members of the LBNL-SNL-UC Berkeley Co-Design and Integration of nano-sensors on CMOS collaboration for useful discussions. This material is based upon work supported by the U.S. Department of Energy, Office of Science, for support of  microelectronics research. Sandia National Laboratories is a multimission laboratory managed and operated by National Technology and Engineering Solutions of Sandia, LLC., a wholly owned subsidiary of Honeywell International, Inc., for the U.S. Department of Energy's National Nuclear Security Administration under contract DE-NA-0003525.  

\section*{Author contributions}

S.Y., M.S., and F.L. conceived the idea. S.Y. developed the numerical code for the simulations and performed the simulations. S.Y., M.S., and F.L. prepared the manuscript. F.L. supervised the project.

\section*{Competing interests}

The Authors declare no Competing Financial or Non-Financial Interests

\bibliography{stoch_draft}

\begin{thebibliography}{41}%
\makeatletter
\providecommand \@ifxundefined [1]{%
 \@ifx{#1\undefined}
}%
\providecommand \@ifnum [1]{%
 \ifnum #1\expandafter \@firstoftwo
 \else \expandafter \@secondoftwo
 \fi
}%
\providecommand \@ifx [1]{%
 \ifx #1\expandafter \@firstoftwo
 \else \expandafter \@secondoftwo
 \fi
}%
\providecommand \natexlab [1]{#1}%
\providecommand \enquote  [1]{``#1''}%
\providecommand \bibnamefont  [1]{#1}%
\providecommand \bibfnamefont [1]{#1}%
\providecommand \citenamefont [1]{#1}%
\providecommand \href@noop [0]{\@secondoftwo}%
\providecommand \href [0]{\begingroup \@sanitize@url \@href}%
\providecommand \@href[1]{\@@startlink{#1}\@@href}%
\providecommand \@@href[1]{\endgroup#1\@@endlink}%
\providecommand \@sanitize@url [0]{\catcode `\\12\catcode `\$12\catcode
  `\&12\catcode `\#12\catcode `\^12\catcode `\_12\catcode `\%12\relax}%
\providecommand \@@startlink[1]{}%
\providecommand \@@endlink[0]{}%
\providecommand \url  [0]{\begingroup\@sanitize@url \@url }%
\providecommand \@url [1]{\endgroup\@href {#1}{\urlprefix }}%
\providecommand \urlprefix  [0]{URL }%
\providecommand \Eprint [0]{\href }%
\providecommand \doibase [0]{http://dx.doi.org/}%
\providecommand \selectlanguage [0]{\@gobble}%
\providecommand \bibinfo  [0]{\@secondoftwo}%
\providecommand \bibfield  [0]{\@secondoftwo}%
\providecommand \translation [1]{[#1]}%
\providecommand \BibitemOpen [0]{}%
\providecommand \bibitemStop [0]{}%
\providecommand \bibitemNoStop [0]{.\EOS\space}%
\providecommand \EOS [0]{\spacefactor3000\relax}%
\providecommand \BibitemShut  [1]{\csname bibitem#1\endcsname}%
\let\auto@bib@innerbib\@empty
\bibitem [{\citenamefont {Chunnilall}\ \emph {et~al.}(2014)\citenamefont
  {Chunnilall}, \citenamefont {Degiovanni}, \citenamefont {K\u{u}ck},
  \citenamefont {M\u{u}ller},\ and\ \citenamefont
  {Sinclair}}]{Chunnilall:2014}%
  \BibitemOpen
  \bibfield  {author} {\bibinfo {author} {\bibfnamefont {C.~J.}\ \bibnamefont
  {Chunnilall}}, \bibinfo {author} {\bibfnamefont {I.~P.}\ \bibnamefont
  {Degiovanni}}, \bibinfo {author} {\bibfnamefont {S.}~\bibnamefont
  {K\u{u}ck}}, \bibinfo {author} {\bibfnamefont {I.}~\bibnamefont
  {M\u{u}ller}}, \ and\ \bibinfo {author} {\bibfnamefont {A.~G.}\ \bibnamefont
  {Sinclair}},\ }\href {\doibase 10.1117/1.OE.53.8.081910} {\bibfield
  {journal} {\bibinfo  {journal} {Opt. Eng.}\ }\textbf {\bibinfo {volume}
  {53}},\ \bibinfo {pages} {1,17} (\bibinfo {year} {2014})}\BibitemShut
  {NoStop}%
\bibitem [{\citenamefont {Bienfang}\ \emph {et~al.}(2004)\citenamefont
  {Bienfang}, \citenamefont {Gross}, \citenamefont {Mink}, \citenamefont
  {Hershman}, \citenamefont {Nakassis}, \citenamefont {Tang}, \citenamefont
  {Lu}, \citenamefont {Su}, \citenamefont {Clark}, \citenamefont {Williams},
  \citenamefont {Hagley},\ and\ \citenamefont {Wen}}]{Bienfang:2004ij}%
  \BibitemOpen
  \bibfield  {author} {\bibinfo {author} {\bibfnamefont {J.~C.}\ \bibnamefont
  {Bienfang}}, \bibinfo {author} {\bibfnamefont {A.~J.}\ \bibnamefont {Gross}},
  \bibinfo {author} {\bibfnamefont {A.}~\bibnamefont {Mink}}, \bibinfo {author}
  {\bibfnamefont {B.~J.}\ \bibnamefont {Hershman}}, \bibinfo {author}
  {\bibfnamefont {A.}~\bibnamefont {Nakassis}}, \bibinfo {author}
  {\bibfnamefont {X.}~\bibnamefont {Tang}}, \bibinfo {author} {\bibfnamefont
  {R.}~\bibnamefont {Lu}}, \bibinfo {author} {\bibfnamefont {D.~H.}\
  \bibnamefont {Su}}, \bibinfo {author} {\bibfnamefont {C.~W.}\ \bibnamefont
  {Clark}}, \bibinfo {author} {\bibfnamefont {C.~J.}\ \bibnamefont {Williams}},
  \bibinfo {author} {\bibfnamefont {E.~W.}\ \bibnamefont {Hagley}}, \ and\
  \bibinfo {author} {\bibfnamefont {J.}~\bibnamefont {Wen}},\ }\href@noop {}
  {\bibfield  {journal} {\bibinfo  {journal} {Opt. Express}\ }\textbf {\bibinfo
  {volume} {12}},\ \bibinfo {pages} {2011} (\bibinfo {year}
  {2004})}\BibitemShut {NoStop}%
\bibitem [{\citenamefont {Woodson}\ \emph {et~al.}(2016)\citenamefont
  {Woodson}, \citenamefont {Ren}, \citenamefont {Maddox}, \citenamefont {Chen},
  \citenamefont {Bank},\ and\ \citenamefont {Campbell}}]{Woodson:2016cx}%
  \BibitemOpen
  \bibfield  {author} {\bibinfo {author} {\bibfnamefont {M.~E.}\ \bibnamefont
  {Woodson}}, \bibinfo {author} {\bibfnamefont {M.}~\bibnamefont {Ren}},
  \bibinfo {author} {\bibfnamefont {S.~J.}\ \bibnamefont {Maddox}}, \bibinfo
  {author} {\bibfnamefont {Y.}~\bibnamefont {Chen}}, \bibinfo {author}
  {\bibfnamefont {S.~R.}\ \bibnamefont {Bank}}, \ and\ \bibinfo {author}
  {\bibfnamefont {J.~C.}\ \bibnamefont {Campbell}},\ }\href@noop {} {\bibfield
  {journal} {\bibinfo  {journal} {Appl. Phys. Lett.}\ }\textbf {\bibinfo
  {volume} {108}},\ \bibinfo {pages} {081102} (\bibinfo {year}
  {2016})}\BibitemShut {NoStop}%
\bibitem [{\citenamefont {Pernice}\ \emph {et~al.}(2012)\citenamefont
  {Pernice}, \citenamefont {Schuck}, \citenamefont {Minaeva}, \citenamefont
  {Li}, \citenamefont {Goltsman}, \citenamefont {Sergienko},\ and\
  \citenamefont {Tang}}]{Pernice:2012bc}%
  \BibitemOpen
  \bibfield  {author} {\bibinfo {author} {\bibfnamefont {W.~H.~P.}\
  \bibnamefont {Pernice}}, \bibinfo {author} {\bibfnamefont {C.}~\bibnamefont
  {Schuck}}, \bibinfo {author} {\bibfnamefont {O.}~\bibnamefont {Minaeva}},
  \bibinfo {author} {\bibfnamefont {M.}~\bibnamefont {Li}}, \bibinfo {author}
  {\bibfnamefont {G.~N.}\ \bibnamefont {Goltsman}}, \bibinfo {author}
  {\bibfnamefont {A.~V.}\ \bibnamefont {Sergienko}}, \ and\ \bibinfo {author}
  {\bibfnamefont {H.~X.}\ \bibnamefont {Tang}},\ }\href@noop {} {\bibfield
  {journal} {\bibinfo  {journal} {Nature}\ }\textbf {\bibinfo {volume} {3}},\
  \bibinfo {pages} {1325} (\bibinfo {year} {2012})}\BibitemShut {NoStop}%
\bibitem [{\citenamefont {Marsili}\ \emph {et~al.}(2012)\citenamefont
  {Marsili}, \citenamefont {Bellei}, \citenamefont {Najafi}, \citenamefont
  {Dane}, \citenamefont {Dauler}, \citenamefont {Molnar},\ and\ \citenamefont
  {Berggren}}]{Marsili:2012ib}%
  \BibitemOpen
  \bibfield  {author} {\bibinfo {author} {\bibfnamefont {F.}~\bibnamefont
  {Marsili}}, \bibinfo {author} {\bibfnamefont {F.}~\bibnamefont {Bellei}},
  \bibinfo {author} {\bibfnamefont {F.}~\bibnamefont {Najafi}}, \bibinfo
  {author} {\bibfnamefont {A.~E.}\ \bibnamefont {Dane}}, \bibinfo {author}
  {\bibfnamefont {E.~A.}\ \bibnamefont {Dauler}}, \bibinfo {author}
  {\bibfnamefont {R.~J.}\ \bibnamefont {Molnar}}, \ and\ \bibinfo {author}
  {\bibfnamefont {K.~K.}\ \bibnamefont {Berggren}},\ }\href@noop {} {\bibfield
  {journal} {\bibinfo  {journal} {Nano. Lett.}\ }\textbf {\bibinfo {volume}
  {12}},\ \bibinfo {pages} {4799} (\bibinfo {year} {2012})}\BibitemShut
  {NoStop}%
\bibitem [{\citenamefont {Marsili}\ \emph {et~al.}(2013)\citenamefont
  {Marsili}, \citenamefont {Verma}, \citenamefont {Stern}, \citenamefont
  {Harrington}, \citenamefont {Lita}, \citenamefont {Gerrits}, \citenamefont
  {Vayshenker}, \citenamefont {Baek}, \citenamefont {Shaw}, \citenamefont
  {Mirin},\ and\ \citenamefont {Nam}}]{Marsili:2013th}%
  \BibitemOpen
  \bibfield  {author} {\bibinfo {author} {\bibfnamefont {F.}~\bibnamefont
  {Marsili}}, \bibinfo {author} {\bibfnamefont {V.~B.}\ \bibnamefont {Verma}},
  \bibinfo {author} {\bibfnamefont {J.~A.}\ \bibnamefont {Stern}}, \bibinfo
  {author} {\bibfnamefont {S.}~\bibnamefont {Harrington}}, \bibinfo {author}
  {\bibfnamefont {A.~E.}\ \bibnamefont {Lita}}, \bibinfo {author}
  {\bibfnamefont {T.}~\bibnamefont {Gerrits}}, \bibinfo {author} {\bibfnamefont
  {I.}~\bibnamefont {Vayshenker}}, \bibinfo {author} {\bibfnamefont
  {B.}~\bibnamefont {Baek}}, \bibinfo {author} {\bibfnamefont {M.~D.}\
  \bibnamefont {Shaw}}, \bibinfo {author} {\bibfnamefont {R.~P.}\ \bibnamefont
  {Mirin}}, \ and\ \bibinfo {author} {\bibfnamefont {S.~W.}\ \bibnamefont
  {Nam}},\ }\href@noop {} {\bibfield  {journal} {\bibinfo  {journal} {Nat.
  Photonics}\ }\textbf {\bibinfo {volume} {7}},\ \bibinfo {pages} {210}
  (\bibinfo {year} {2013})}\BibitemShut {NoStop}%
\bibitem [{\citenamefont {Eisaman}\ \emph {et~al.}(2011)\citenamefont
  {Eisaman}, \citenamefont {Fan}, \citenamefont {Migdall},\ and\ \citenamefont
  {Polyakov}}]{Eisaman:2011cc}%
  \BibitemOpen
  \bibfield  {author} {\bibinfo {author} {\bibfnamefont {M.~D.}\ \bibnamefont
  {Eisaman}}, \bibinfo {author} {\bibfnamefont {J.}~\bibnamefont {Fan}},
  \bibinfo {author} {\bibfnamefont {A.}~\bibnamefont {Migdall}}, \ and\
  \bibinfo {author} {\bibfnamefont {S.~V.}\ \bibnamefont {Polyakov}},\
  }\href@noop {} {\bibfield  {journal} {\bibinfo  {journal} {Review of
  Scientific Instruments}\ }\textbf {\bibinfo {volume} {82}},\ \bibinfo {pages}
  {071101} (\bibinfo {year} {2011})}\BibitemShut {NoStop}%
\bibitem [{\citenamefont {Hadfield}(2009)}]{Hadfield:2009}%
  \BibitemOpen
  \bibfield  {author} {\bibinfo {author} {\bibfnamefont {R.~H.}\ \bibnamefont
  {Hadfield}},\ }\href@noop {} {\bibfield  {journal} {\bibinfo  {journal} {Nat.
  Photonics}\ }\textbf {\bibinfo {volume} {3}},\ \bibinfo {pages} {696}
  (\bibinfo {year} {2009})}\BibitemShut {NoStop}%
\bibitem [{\citenamefont {Rosenberg}\ \emph {et~al.}(2005)\citenamefont
  {Rosenberg}, \citenamefont {Lita}, \citenamefont {Miller},\ and\
  \citenamefont {Nam}}]{Rosenberg:2005}%
  \BibitemOpen
  \bibfield  {author} {\bibinfo {author} {\bibfnamefont {D.}~\bibnamefont
  {Rosenberg}}, \bibinfo {author} {\bibfnamefont {A.~E.}\ \bibnamefont {Lita}},
  \bibinfo {author} {\bibfnamefont {A.~J.}\ \bibnamefont {Miller}}, \ and\
  \bibinfo {author} {\bibfnamefont {S.~W.}\ \bibnamefont {Nam}},\ }\href
  {\doibase 10.1103/PhysRevA.71.061803} {\bibfield  {journal} {\bibinfo
  {journal} {Phys. Rev. A}\ }\textbf {\bibinfo {volume} {71}},\ \bibinfo
  {pages} {061803} (\bibinfo {year} {2005})}\BibitemShut {NoStop}%
\bibitem [{\citenamefont {Kardynal}\ \emph {et~al.}(2008)\citenamefont
  {Kardynal}, \citenamefont {Yuan},\ and\ \citenamefont
  {Shields}}]{Kardynal:2008}%
  \BibitemOpen
  \bibfield  {author} {\bibinfo {author} {\bibfnamefont {B.}~\bibnamefont
  {Kardynal}}, \bibinfo {author} {\bibfnamefont {Z.~L.}\ \bibnamefont {Yuan}},
  \ and\ \bibinfo {author} {\bibfnamefont {A.~J.}\ \bibnamefont {Shields}},\
  }\href@noop {} {\bibfield  {journal} {\bibinfo  {journal} {Nature Photonics}\
  }\textbf {\bibinfo {volume} {2}},\ \bibinfo {pages} {425} (\bibinfo {year}
  {2008})}\BibitemShut {NoStop}%
\bibitem [{\citenamefont {Young}\ \emph {et~al.}(2020)\citenamefont {Young},
  \citenamefont {Sarovar},\ and\ \citenamefont {L\'{e}onard}}]{Young:2020}%
  \BibitemOpen
  \bibfield  {author} {\bibinfo {author} {\bibfnamefont {S.~M.}\ \bibnamefont
  {Young}}, \bibinfo {author} {\bibfnamefont {M.}~\bibnamefont {Sarovar}}, \
  and\ \bibinfo {author} {\bibfnamefont {F.}~\bibnamefont {L\'{e}onard}},\
  }\href {\doibase 10.1021/acsphotonics.9b01754} {\bibfield  {journal}
  {\bibinfo  {journal} {ACS Photonics}\ }\textbf {\bibinfo {volume} {7}},\
  \bibinfo {pages} {821} (\bibinfo {year} {2020})}\BibitemShut {NoStop}%
\bibitem [{\citenamefont {Griffiths}\ \emph {et~al.}(2018)\citenamefont
  {Griffiths}, \citenamefont {Chen}, \citenamefont {Herrnsdorf}, \citenamefont
  {Li}, \citenamefont {Henderson}, \citenamefont {Strain},\ and\ \citenamefont
  {Dawson}}]{Griffiths:2018}%
  \BibitemOpen
  \bibfield  {author} {\bibinfo {author} {\bibfnamefont {A.~D.}\ \bibnamefont
  {Griffiths}}, \bibinfo {author} {\bibfnamefont {H.}~\bibnamefont {Chen}},
  \bibinfo {author} {\bibfnamefont {J.}~\bibnamefont {Herrnsdorf}}, \bibinfo
  {author} {\bibfnamefont {D.}~\bibnamefont {Li}}, \bibinfo {author}
  {\bibfnamefont {R.~K.}\ \bibnamefont {Henderson}}, \bibinfo {author}
  {\bibfnamefont {M.~J.}\ \bibnamefont {Strain}}, \ and\ \bibinfo {author}
  {\bibfnamefont {M.~D.}\ \bibnamefont {Dawson}},\ }\href {\doibase
  10.1109/BICOP.2018.8658323} {\bibfield  {journal} {\bibinfo  {journal} {2018
  IEEE British and Irish Conference on Optics and Photonics (BICOP)}\ ,\
  \bibinfo {pages} {1}} (\bibinfo {year} {2018})}\BibitemShut {NoStop}%
\bibitem [{\citenamefont {Collaboration}(2016)}]{Aghamousa}%
  \BibitemOpen
  \bibfield  {author} {\bibinfo {author} {\bibfnamefont {D.}~\bibnamefont
  {Collaboration}},\ }\href {\doibase 10.48550/ARXIV.1611.00037} {\bibfield
  {journal} {\bibinfo  {journal} {arXiv}\ ,\ \bibinfo {pages}
  {ARXIV.1611.00037}} (\bibinfo {year} {2016})}\BibitemShut {NoStop}%
\bibitem [{\citenamefont {Niwa}\ \emph {et~al.}(2021)\citenamefont {Niwa},
  \citenamefont {Hattori},\ and\ \citenamefont {Fukuda}}]{Niwa:2021}%
  \BibitemOpen
  \bibfield  {author} {\bibinfo {author} {\bibfnamefont {K.}~\bibnamefont
  {Niwa}}, \bibinfo {author} {\bibfnamefont {K.}~\bibnamefont {Hattori}}, \
  and\ \bibinfo {author} {\bibfnamefont {D.}~\bibnamefont {Fukuda}},\
  }\href@noop {} {\bibfield  {journal} {\bibinfo  {journal} {Frontiers in
  Bioengineering and Biotechnology}\ }\textbf {\bibinfo {volume} {9}} (\bibinfo
  {year} {2021})}\BibitemShut {NoStop}%
\bibitem [{\citenamefont {Judd}(1953)}]{Judd1953}%
  \BibitemOpen
  \bibfield  {author} {\bibinfo {author} {\bibfnamefont {D.~B.}\ \bibnamefont
  {Judd}},\ }\href {\doibase 10.1002/jps.3030421221} {\bibfield  {journal}
  {\bibinfo  {journal} {J. Am. Pharm. Assoc. Sci.}\ }\textbf {\bibinfo {volume}
  {42}},\ \bibinfo {pages} {757} (\bibinfo {year} {1953})}\BibitemShut
  {NoStop}%
\bibitem [{\citenamefont {Kahl}\ \emph {et~al.}(2017)\citenamefont {Kahl},
  \citenamefont {Ferrari}, \citenamefont {Kovalyuk}, \citenamefont {Vetter},
  \citenamefont {Lewes-Malandrakis}, \citenamefont {Nebel}, \citenamefont
  {Korneev}, \citenamefont {Goltsman},\ and\ \citenamefont
  {Pernice}}]{Kahl:17}%
  \BibitemOpen
  \bibfield  {author} {\bibinfo {author} {\bibfnamefont {O.}~\bibnamefont
  {Kahl}}, \bibinfo {author} {\bibfnamefont {S.}~\bibnamefont {Ferrari}},
  \bibinfo {author} {\bibfnamefont {V.}~\bibnamefont {Kovalyuk}}, \bibinfo
  {author} {\bibfnamefont {A.}~\bibnamefont {Vetter}}, \bibinfo {author}
  {\bibfnamefont {G.}~\bibnamefont {Lewes-Malandrakis}}, \bibinfo {author}
  {\bibfnamefont {C.}~\bibnamefont {Nebel}}, \bibinfo {author} {\bibfnamefont
  {A.}~\bibnamefont {Korneev}}, \bibinfo {author} {\bibfnamefont
  {G.}~\bibnamefont {Goltsman}}, \ and\ \bibinfo {author} {\bibfnamefont
  {W.}~\bibnamefont {Pernice}},\ }\href {\doibase 10.1364/OPTICA.4.000557}
  {\bibfield  {journal} {\bibinfo  {journal} {Optica}\ }\textbf {\bibinfo
  {volume} {4}},\ \bibinfo {pages} {557} (\bibinfo {year} {2017})}\BibitemShut
  {NoStop}%
\bibitem [{\citenamefont {Cheng}\ \emph {et~al.}(2019)\citenamefont {Cheng},
  \citenamefont {Zou}, \citenamefont {Guo}, \citenamefont {Wang}, \citenamefont
  {Han},\ and\ \citenamefont {Tang}}]{Cheng:2019}%
  \BibitemOpen
  \bibfield  {author} {\bibinfo {author} {\bibfnamefont {R.}~\bibnamefont
  {Cheng}}, \bibinfo {author} {\bibfnamefont {C.-L.}\ \bibnamefont {Zou}},
  \bibinfo {author} {\bibfnamefont {X.}~\bibnamefont {Guo}}, \bibinfo {author}
  {\bibfnamefont {S.}~\bibnamefont {Wang}}, \bibinfo {author} {\bibfnamefont
  {X.}~\bibnamefont {Han}}, \ and\ \bibinfo {author} {\bibfnamefont {H.~X.}\
  \bibnamefont {Tang}},\ }\href@noop {} {\bibfield  {journal} {\bibinfo
  {journal} {Nat. Comm.}\ }\textbf {\bibinfo {volume} {10}},\ \bibinfo {pages}
  {4104} (\bibinfo {year} {2019})}\BibitemShut {NoStop}%
\bibitem [{\citenamefont {Ma}\ \emph {et~al.}(2017)\citenamefont {Ma},
  \citenamefont {Slattery},\ and\ \citenamefont {Tang}}]{Ma:17}%
  \BibitemOpen
  \bibfield  {author} {\bibinfo {author} {\bibfnamefont {L.}~\bibnamefont
  {Ma}}, \bibinfo {author} {\bibfnamefont {O.}~\bibnamefont {Slattery}}, \ and\
  \bibinfo {author} {\bibfnamefont {X.}~\bibnamefont {Tang}},\ }\href@noop {}
  {\bibfield  {journal} {\bibinfo  {journal} {Opt. Express}\ }\textbf {\bibinfo
  {volume} {25}},\ \bibinfo {pages} {28898} (\bibinfo {year}
  {2017})}\BibitemShut {NoStop}%
\bibitem [{\citenamefont {Cabrera}\ \emph {et~al.}(1998)\citenamefont
  {Cabrera}, \citenamefont {Clarke}, \citenamefont {Colling}, \citenamefont
  {Miller}, \citenamefont {Nam},\ and\ \citenamefont {Romani}}]{Cabrera1998}%
  \BibitemOpen
  \bibfield  {author} {\bibinfo {author} {\bibfnamefont {B.}~\bibnamefont
  {Cabrera}}, \bibinfo {author} {\bibfnamefont {R.~M.}\ \bibnamefont {Clarke}},
  \bibinfo {author} {\bibfnamefont {P.}~\bibnamefont {Colling}}, \bibinfo
  {author} {\bibfnamefont {A.~J.}\ \bibnamefont {Miller}}, \bibinfo {author}
  {\bibfnamefont {S.}~\bibnamefont {Nam}}, \ and\ \bibinfo {author}
  {\bibfnamefont {R.~W.}\ \bibnamefont {Romani}},\ }\href@noop {} {\bibfield
  {journal} {\bibinfo  {journal} {Applied Physics Letters}\ }\textbf {\bibinfo
  {volume} {73}},\ \bibinfo {pages} {735} (\bibinfo {year} {1998})}\BibitemShut
  {NoStop}%
\bibitem [{\citenamefont {Fukuda}\ \emph {et~al.}(2011)\citenamefont {Fukuda},
  \citenamefont {Fujii}, \citenamefont {Numata}, \citenamefont {Amemiya},
  \citenamefont {Yoshizawa}, \citenamefont {Tsuchida}, \citenamefont {Fujino},
  \citenamefont {Ishii}, \citenamefont {Itatani}, \citenamefont {Inoue},\ and\
  \citenamefont {Zama}}]{Fukuda_2011}%
  \BibitemOpen
  \bibfield  {author} {\bibinfo {author} {\bibfnamefont {D.}~\bibnamefont
  {Fukuda}}, \bibinfo {author} {\bibfnamefont {G.}~\bibnamefont {Fujii}},
  \bibinfo {author} {\bibfnamefont {T.}~\bibnamefont {Numata}}, \bibinfo
  {author} {\bibfnamefont {K.}~\bibnamefont {Amemiya}}, \bibinfo {author}
  {\bibfnamefont {A.}~\bibnamefont {Yoshizawa}}, \bibinfo {author}
  {\bibfnamefont {H.}~\bibnamefont {Tsuchida}}, \bibinfo {author}
  {\bibfnamefont {H.}~\bibnamefont {Fujino}}, \bibinfo {author} {\bibfnamefont
  {H.}~\bibnamefont {Ishii}}, \bibinfo {author} {\bibfnamefont
  {T.}~\bibnamefont {Itatani}}, \bibinfo {author} {\bibfnamefont
  {S.}~\bibnamefont {Inoue}}, \ and\ \bibinfo {author} {\bibfnamefont
  {T.}~\bibnamefont {Zama}},\ }\href@noop {} {\bibfield  {journal} {\bibinfo
  {journal} {Opt. Express}\ }\textbf {\bibinfo {volume} {19}},\ \bibinfo
  {pages} {870} (\bibinfo {year} {2011})}\BibitemShut {NoStop}%
\bibitem [{\citenamefont {Hattori}\ \emph {et~al.}(2022)\citenamefont
  {Hattori}, \citenamefont {Konno}, \citenamefont {Miura}, \citenamefont
  {Takasu},\ and\ \citenamefont {Fukuda}}]{Hattori_2022}%
  \BibitemOpen
  \bibfield  {author} {\bibinfo {author} {\bibfnamefont {K.}~\bibnamefont
  {Hattori}}, \bibinfo {author} {\bibfnamefont {T.}~\bibnamefont {Konno}},
  \bibinfo {author} {\bibfnamefont {Y.}~\bibnamefont {Miura}}, \bibinfo
  {author} {\bibfnamefont {S.}~\bibnamefont {Takasu}}, \ and\ \bibinfo {author}
  {\bibfnamefont {D.}~\bibnamefont {Fukuda}},\ }\href {\doibase
  10.1088/1361-6668/ac7e7b} {\bibfield  {journal} {\bibinfo  {journal}
  {Superconductor Science and Technology}\ }\textbf {\bibinfo {volume} {35}},\
  \bibinfo {pages} {095002} (\bibinfo {year} {2022})}\BibitemShut {NoStop}%
\bibitem [{\citenamefont {Young}\ \emph
  {et~al.}(2018{\natexlab{a}})\citenamefont {Young}, \citenamefont {Sarovar},\
  and\ \citenamefont {L\'eonard}}]{Young:2018b}%
  \BibitemOpen
  \bibfield  {author} {\bibinfo {author} {\bibfnamefont {S.~M.}\ \bibnamefont
  {Young}}, \bibinfo {author} {\bibfnamefont {M.}~\bibnamefont {Sarovar}}, \
  and\ \bibinfo {author} {\bibfnamefont {F.}~\bibnamefont {L\'eonard}},\ }\href
  {\doibase 10.1103/PhysRevA.98.063835} {\bibfield  {journal} {\bibinfo
  {journal} {Phys. Rev. A}\ }\textbf {\bibinfo {volume} {98}},\ \bibinfo
  {pages} {063835} (\bibinfo {year} {2018}{\natexlab{a}})}\BibitemShut
  {NoStop}%
\bibitem [{\citenamefont {Bienaim{\'{e}}}\ \emph {et~al.}(2012)\citenamefont
  {Bienaim{\'{e}}}, \citenamefont {Bachelard}, \citenamefont {Piovella},\ and\
  \citenamefont {Kaiser}}]{Bienaime2012}%
  \BibitemOpen
  \bibfield  {author} {\bibinfo {author} {\bibfnamefont {T.}~\bibnamefont
  {Bienaim{\'{e}}}}, \bibinfo {author} {\bibfnamefont {R.}~\bibnamefont
  {Bachelard}}, \bibinfo {author} {\bibfnamefont {N.}~\bibnamefont {Piovella}},
  \ and\ \bibinfo {author} {\bibfnamefont {R.}~\bibnamefont {Kaiser}},\ }\href
  {\doibase 10.1002/prop.201200089} {\bibfield  {journal} {\bibinfo  {journal}
  {Fortschritte der Physik}\ }\textbf {\bibinfo {volume} {61}},\ \bibinfo
  {pages} {377} (\bibinfo {year} {2012})}\BibitemShut {NoStop}%
\bibitem [{\citenamefont {Reitz}\ \emph {et~al.}(2022)\citenamefont {Reitz},
  \citenamefont {Sommer},\ and\ \citenamefont {Genes}}]{Reitz2022}%
  \BibitemOpen
  \bibfield  {author} {\bibinfo {author} {\bibfnamefont {M.}~\bibnamefont
  {Reitz}}, \bibinfo {author} {\bibfnamefont {C.}~\bibnamefont {Sommer}}, \
  and\ \bibinfo {author} {\bibfnamefont {C.}~\bibnamefont {Genes}},\ }\href
  {\doibase 10.1103/PRXQuantum.3.010201} {\bibfield  {journal} {\bibinfo
  {journal} {PRX Quantum}\ }\textbf {\bibinfo {volume} {3}},\ \bibinfo {pages}
  {010201} (\bibinfo {year} {2022})}\BibitemShut {NoStop}%
\bibitem [{\citenamefont {Baragiola}\ \emph {et~al.}(2012)\citenamefont
  {Baragiola}, \citenamefont {Cook}, \citenamefont {Branczyk},\ and\
  \citenamefont {Combes}}]{Baragiola:2012cs}%
  \BibitemOpen
  \bibfield  {author} {\bibinfo {author} {\bibfnamefont {B.~Q.}\ \bibnamefont
  {Baragiola}}, \bibinfo {author} {\bibfnamefont {R.~L.}\ \bibnamefont {Cook}},
  \bibinfo {author} {\bibfnamefont {A.~M.}\ \bibnamefont {Branczyk}}, \ and\
  \bibinfo {author} {\bibfnamefont {J.}~\bibnamefont {Combes}},\ }\href@noop {}
  {\bibfield  {journal} {\bibinfo  {journal} {Phys. Rev. A}\ }\textbf {\bibinfo
  {volume} {86}},\ \bibinfo {pages} {013811} (\bibinfo {year}
  {2012})}\BibitemShut {NoStop}%
\bibitem [{\citenamefont {Young}\ \emph
  {et~al.}(2018{\natexlab{b}})\citenamefont {Young}, \citenamefont {Sarovar},\
  and\ \citenamefont {L\'eonard}}]{Young:2018}%
  \BibitemOpen
  \bibfield  {author} {\bibinfo {author} {\bibfnamefont {S.~M.}\ \bibnamefont
  {Young}}, \bibinfo {author} {\bibfnamefont {M.}~\bibnamefont {Sarovar}}, \
  and\ \bibinfo {author} {\bibfnamefont {F.}~\bibnamefont {L\'eonard}},\ }\href
  {\doibase 10.1103/PhysRevA.97.033836} {\bibfield  {journal} {\bibinfo
  {journal} {Phys. Rev. A}\ }\textbf {\bibinfo {volume} {97}},\ \bibinfo
  {pages} {033836} (\bibinfo {year} {2018}{\natexlab{b}})}\BibitemShut
  {NoStop}%
\bibitem [{\citenamefont {Mattiotti}\ \emph {et~al.}(2022)\citenamefont
  {Mattiotti}, \citenamefont {Sarovar}, \citenamefont {Giusteri}, \citenamefont
  {Borgonovi},\ and\ \citenamefont {Celardo}}]{Mattiotti_2022}%
  \BibitemOpen
  \bibfield  {author} {\bibinfo {author} {\bibfnamefont {F.}~\bibnamefont
  {Mattiotti}}, \bibinfo {author} {\bibfnamefont {M.}~\bibnamefont {Sarovar}},
  \bibinfo {author} {\bibfnamefont {G.~G.}\ \bibnamefont {Giusteri}}, \bibinfo
  {author} {\bibfnamefont {F.}~\bibnamefont {Borgonovi}}, \ and\ \bibinfo
  {author} {\bibfnamefont {G.~L.}\ \bibnamefont {Celardo}},\ }\href {\doibase
  10.1088/1367-2630/ac4127} {\bibfield  {journal} {\bibinfo  {journal} {New
  Journal of Physics}\ }\textbf {\bibinfo {volume} {24}},\ \bibinfo {pages}
  {013027} (\bibinfo {year} {2022})}\BibitemShut {NoStop}%
\bibitem [{\citenamefont {Byrd}\ \emph {et~al.}(1995)\citenamefont {Byrd},
  \citenamefont {Lu}, \citenamefont {Nocedal},\ and\ \citenamefont
  {Zhu}}]{Byrd95}%
  \BibitemOpen
  \bibfield  {author} {\bibinfo {author} {\bibfnamefont {R.~H.}\ \bibnamefont
  {Byrd}}, \bibinfo {author} {\bibfnamefont {P.}~\bibnamefont {Lu}}, \bibinfo
  {author} {\bibfnamefont {J.}~\bibnamefont {Nocedal}}, \ and\ \bibinfo
  {author} {\bibfnamefont {C.}~\bibnamefont {Zhu}},\ }\href {\doibase
  10.1137/0916069} {\bibfield  {journal} {\bibinfo  {journal} {{SIAM} Journal
  on Scientific Computing}\ }\textbf {\bibinfo {volume} {16}},\ \bibinfo
  {pages} {1190} (\bibinfo {year} {1995})}\BibitemShut {NoStop}%
\bibitem [{\citenamefont {Bergemann}\ and\ \citenamefont
  {L\'eonard}(2018)}]{bergemann}%
  \BibitemOpen
  \bibfield  {author} {\bibinfo {author} {\bibfnamefont {K.}~\bibnamefont
  {Bergemann}}\ and\ \bibinfo {author} {\bibfnamefont {F.}~\bibnamefont
  {L\'eonard}},\ }\href {\doibase 10.1002/smll.201802806} {\bibfield  {journal}
  {\bibinfo  {journal} {Small}\ }\textbf {\bibinfo {volume} {14}},\ \bibinfo
  {pages} {1802806} (\bibinfo {year} {2018})}\BibitemShut {NoStop}%
\bibitem [{\citenamefont {Weng}\ \emph {et~al.}(2015)\citenamefont {Weng},
  \citenamefont {An}, \citenamefont {Zhang}, \citenamefont {Chen},
  \citenamefont {Chen}, \citenamefont {Zhu},\ and\ \citenamefont
  {Lu}}]{Weng_2015}%
  \BibitemOpen
  \bibfield  {author} {\bibinfo {author} {\bibfnamefont {Q.}~\bibnamefont
  {Weng}}, \bibinfo {author} {\bibfnamefont {Z.}~\bibnamefont {An}}, \bibinfo
  {author} {\bibfnamefont {B.}~\bibnamefont {Zhang}}, \bibinfo {author}
  {\bibfnamefont {P.}~\bibnamefont {Chen}}, \bibinfo {author} {\bibfnamefont
  {X.}~\bibnamefont {Chen}}, \bibinfo {author} {\bibfnamefont {Z.}~\bibnamefont
  {Zhu}}, \ and\ \bibinfo {author} {\bibfnamefont {W.}~\bibnamefont {Lu}},\
  }\href@noop {} {\bibfield  {journal} {\bibinfo  {journal} {Sci. Rep.}\
  }\textbf {\bibinfo {volume} {5}},\ \bibinfo {pages} {9389} (\bibinfo {year}
  {2015})}\BibitemShut {NoStop}%
\bibitem [{\citenamefont {Ka}\ \emph {et~al.}(2012)\citenamefont {Ka},
  \citenamefont {Le~Borgne}, \citenamefont {Ma},\ and\ \citenamefont
  {El~Khakani}}]{Ka_2012}%
  \BibitemOpen
  \bibfield  {author} {\bibinfo {author} {\bibfnamefont {I.}~\bibnamefont
  {Ka}}, \bibinfo {author} {\bibfnamefont {V.}~\bibnamefont {Le~Borgne}},
  \bibinfo {author} {\bibfnamefont {D.}~\bibnamefont {Ma}}, \ and\ \bibinfo
  {author} {\bibfnamefont {M.~A.}\ \bibnamefont {El~Khakani}},\ }\href@noop {}
  {\bibfield  {journal} {\bibinfo  {journal} {Advanced Materials}\ }\textbf
  {\bibinfo {volume} {24}},\ \bibinfo {pages} {6288} (\bibinfo {year}
  {2012})}\BibitemShut {NoStop}%
\bibitem [{\citenamefont {Ye}\ \emph {et~al.}(2021)\citenamefont {Ye},
  \citenamefont {Xu}, \citenamefont {Paghi}, \citenamefont {Bamford},
  \citenamefont {Horrocks}, \citenamefont {Houlton}, \citenamefont {Barillaro},
  \citenamefont {Dimitrov},\ and\ \citenamefont {Palma}}]{Ye_2021}%
  \BibitemOpen
  \bibfield  {author} {\bibinfo {author} {\bibfnamefont {Q.}~\bibnamefont
  {Ye}}, \bibinfo {author} {\bibfnamefont {X.}~\bibnamefont {Xu}}, \bibinfo
  {author} {\bibfnamefont {A.}~\bibnamefont {Paghi}}, \bibinfo {author}
  {\bibfnamefont {T.}~\bibnamefont {Bamford}}, \bibinfo {author} {\bibfnamefont
  {B.~R.}\ \bibnamefont {Horrocks}}, \bibinfo {author} {\bibfnamefont
  {A.}~\bibnamefont {Houlton}}, \bibinfo {author} {\bibfnamefont
  {G.}~\bibnamefont {Barillaro}}, \bibinfo {author} {\bibfnamefont
  {S.}~\bibnamefont {Dimitrov}}, \ and\ \bibinfo {author} {\bibfnamefont
  {M.}~\bibnamefont {Palma}},\ }\href@noop {} {\bibfield  {journal} {\bibinfo
  {journal} {Advanced Functional Materials}\ }\textbf {\bibinfo {volume}
  {31}},\ \bibinfo {pages} {2105719} (\bibinfo {year} {2021})}\BibitemShut
  {NoStop}%
\bibitem [{\citenamefont {Spataru}\ and\ \citenamefont
  {L\'eonard}(2019)}]{Spataru_2019}%
  \BibitemOpen
  \bibfield  {author} {\bibinfo {author} {\bibfnamefont {C.~D.}\ \bibnamefont
  {Spataru}}\ and\ \bibinfo {author} {\bibfnamefont {F.}~\bibnamefont
  {L\'eonard}},\ }\href@noop {} {\bibfield  {journal} {\bibinfo  {journal}
  {Phys. Rev. Research}\ }\textbf {\bibinfo {volume} {1}},\ \bibinfo {pages}
  {013018} (\bibinfo {year} {2019})}\BibitemShut {NoStop}%
\bibitem [{\citenamefont {L\'eonard}\ \emph {et~al.}(2019)\citenamefont
  {L\'eonard}, \citenamefont {Foster},\ and\ \citenamefont
  {Spataru}}]{Leonard_2019}%
  \BibitemOpen
  \bibfield  {author} {\bibinfo {author} {\bibfnamefont {F.}~\bibnamefont
  {L\'eonard}}, \bibinfo {author} {\bibfnamefont {M.}~\bibnamefont {Foster}}, \
  and\ \bibinfo {author} {\bibfnamefont {C.}~\bibnamefont {Spataru}},\
  }\href@noop {} {\bibfield  {journal} {\bibinfo  {journal} {Sci. Rep.}\ ,\
  \bibinfo {pages} {3268}} (\bibinfo {year} {2019})}\BibitemShut {NoStop}%
\bibitem [{\citenamefont {Cui}\ \emph {et~al.}(2019)\citenamefont {Cui},
  \citenamefont {Panfil}, \citenamefont {Koley}, \citenamefont {Shamalia},
  \citenamefont {Waiskopf}, \citenamefont {Remennik}, \citenamefont {Popov},
  \citenamefont {Oded},\ and\ \citenamefont {Banin}}]{Cui2019}%
  \BibitemOpen
  \bibfield  {author} {\bibinfo {author} {\bibfnamefont {J.}~\bibnamefont
  {Cui}}, \bibinfo {author} {\bibfnamefont {Y.~E.}\ \bibnamefont {Panfil}},
  \bibinfo {author} {\bibfnamefont {S.}~\bibnamefont {Koley}}, \bibinfo
  {author} {\bibfnamefont {D.}~\bibnamefont {Shamalia}}, \bibinfo {author}
  {\bibfnamefont {N.}~\bibnamefont {Waiskopf}}, \bibinfo {author}
  {\bibfnamefont {S.}~\bibnamefont {Remennik}}, \bibinfo {author}
  {\bibfnamefont {I.}~\bibnamefont {Popov}}, \bibinfo {author} {\bibfnamefont
  {M.}~\bibnamefont {Oded}}, \ and\ \bibinfo {author} {\bibfnamefont
  {U.}~\bibnamefont {Banin}},\ }\href {\doibase 10.1038/s41467-019-13349-1}
  {\bibfield  {journal} {\bibinfo  {journal} {Nature Communications}\ }\textbf
  {\bibinfo {volume} {10}} (\bibinfo {year} {2019}),\
  10.1038/s41467-019-13349-1}\BibitemShut {NoStop}%
\bibitem [{\citenamefont {Meulenberg}\ \emph {et~al.}(2009)\citenamefont
  {Meulenberg}, \citenamefont {Lee}, \citenamefont {Wolcott}, \citenamefont
  {Zhang}, \citenamefont {Terminello},\ and\ \citenamefont {van
  Buuren}}]{Meulenberg:2009}%
  \BibitemOpen
  \bibfield  {author} {\bibinfo {author} {\bibfnamefont {R.~W.}\ \bibnamefont
  {Meulenberg}}, \bibinfo {author} {\bibfnamefont {J.~R.}\ \bibnamefont {Lee}},
  \bibinfo {author} {\bibfnamefont {A.}~\bibnamefont {Wolcott}}, \bibinfo
  {author} {\bibfnamefont {J.~Z.}\ \bibnamefont {Zhang}}, \bibinfo {author}
  {\bibfnamefont {L.~J.}\ \bibnamefont {Terminello}}, \ and\ \bibinfo {author}
  {\bibfnamefont {T.}~\bibnamefont {van Buuren}},\ }\href@noop {} {\bibfield
  {journal} {\bibinfo  {journal} {ACS Nano}\ }\textbf {\bibinfo {volume} {3}},\
  \bibinfo {pages} {325} (\bibinfo {year} {2009})}\BibitemShut {NoStop}%
\bibitem [{\citenamefont {Rain{\`{o}}}\ \emph {et~al.}(2016)\citenamefont
  {Rain{\`{o}}}, \citenamefont {Nedelcu}, \citenamefont {Protesescu},
  \citenamefont {Bodnarchuk}, \citenamefont {Kovalenko}, \citenamefont
  {Mahrt},\ and\ \citenamefont {St\"{o}ferle}}]{Raino2016}%
  \BibitemOpen
  \bibfield  {author} {\bibinfo {author} {\bibfnamefont {G.}~\bibnamefont
  {Rain{\`{o}}}}, \bibinfo {author} {\bibfnamefont {G.}~\bibnamefont
  {Nedelcu}}, \bibinfo {author} {\bibfnamefont {L.}~\bibnamefont {Protesescu}},
  \bibinfo {author} {\bibfnamefont {M.~I.}\ \bibnamefont {Bodnarchuk}},
  \bibinfo {author} {\bibfnamefont {M.~V.}\ \bibnamefont {Kovalenko}}, \bibinfo
  {author} {\bibfnamefont {R.~F.}\ \bibnamefont {Mahrt}}, \ and\ \bibinfo
  {author} {\bibfnamefont {T.}~\bibnamefont {St\"{o}ferle}},\ }\href {\doibase
  10.1021/acsnano.5b07328} {\bibfield  {journal} {\bibinfo  {journal} {{ACS}
  Nano}\ }\textbf {\bibinfo {volume} {10}},\ \bibinfo {pages} {2485} (\bibinfo
  {year} {2016})}\BibitemShut {NoStop}%
\bibitem [{\citenamefont {Riaz}\ \emph {et~al.}(2019)\citenamefont {Riaz},
  \citenamefont {Alam}, \citenamefont {Selvasundaram}, \citenamefont {Dehm},
  \citenamefont {Hennrich}, \citenamefont {Kappes},\ and\ \citenamefont
  {Krupke}}]{Riaz_2019}%
  \BibitemOpen
  \bibfield  {author} {\bibinfo {author} {\bibfnamefont {A.}~\bibnamefont
  {Riaz}}, \bibinfo {author} {\bibfnamefont {A.}~\bibnamefont {Alam}}, \bibinfo
  {author} {\bibfnamefont {P.~B.}\ \bibnamefont {Selvasundaram}}, \bibinfo
  {author} {\bibfnamefont {S.}~\bibnamefont {Dehm}}, \bibinfo {author}
  {\bibfnamefont {F.}~\bibnamefont {Hennrich}}, \bibinfo {author}
  {\bibfnamefont {M.~M.}\ \bibnamefont {Kappes}}, \ and\ \bibinfo {author}
  {\bibfnamefont {R.}~\bibnamefont {Krupke}},\ }\href@noop {} {\bibfield
  {journal} {\bibinfo  {journal} {Advanced Electronic Materials}\ }\textbf
  {\bibinfo {volume} {5}},\ \bibinfo {pages} {1800265} (\bibinfo {year}
  {2019})}\BibitemShut {NoStop}%
\bibitem [{\citenamefont {Ma}\ \emph {et~al.}(2020)\citenamefont {Ma},
  \citenamefont {Yang}, \citenamefont {Liu}, \citenamefont {Wang},\ and\
  \citenamefont {Peng}}]{Ma_2020}%
  \BibitemOpen
  \bibfield  {author} {\bibinfo {author} {\bibfnamefont {Z.}~\bibnamefont
  {Ma}}, \bibinfo {author} {\bibfnamefont {L.}~\bibnamefont {Yang}}, \bibinfo
  {author} {\bibfnamefont {L.}~\bibnamefont {Liu}}, \bibinfo {author}
  {\bibfnamefont {S.}~\bibnamefont {Wang}}, \ and\ \bibinfo {author}
  {\bibfnamefont {L.-M.}\ \bibnamefont {Peng}},\ }\href@noop {} {\bibfield
  {journal} {\bibinfo  {journal} {ACS Nano}\ }\textbf {\bibinfo {volume}
  {14}},\ \bibinfo {pages} {7191} (\bibinfo {year} {2020})}\BibitemShut
  {NoStop}%
\bibitem [{\citenamefont {Cao}\ \emph {et~al.}(2017)\citenamefont {Cao},
  \citenamefont {Tersoff}, \citenamefont {Farmer}, \citenamefont {Zhu},\ and\
  \citenamefont {Han}}]{Cao_2017}%
  \BibitemOpen
  \bibfield  {author} {\bibinfo {author} {\bibfnamefont {Q.}~\bibnamefont
  {Cao}}, \bibinfo {author} {\bibfnamefont {J.}~\bibnamefont {Tersoff}},
  \bibinfo {author} {\bibfnamefont {D.~B.}\ \bibnamefont {Farmer}}, \bibinfo
  {author} {\bibfnamefont {Y.}~\bibnamefont {Zhu}}, \ and\ \bibinfo {author}
  {\bibfnamefont {S.-J.}\ \bibnamefont {Han}},\ }\href@noop {} {\bibfield
  {journal} {\bibinfo  {journal} {Science}\ }\textbf {\bibinfo {volume}
  {356}},\ \bibinfo {pages} {1369} (\bibinfo {year} {2017})}\BibitemShut
  {NoStop}%
\bibitem [{\citenamefont {Manfrinato}\ \emph {et~al.}(2017)\citenamefont
  {Manfrinato}, \citenamefont {Stein}, \citenamefont {Zhang}, \citenamefont
  {Nam}, \citenamefont {Yager}, \citenamefont {Stach},\ and\ \citenamefont
  {Black}}]{Manfrinato_2017}%
  \BibitemOpen
  \bibfield  {author} {\bibinfo {author} {\bibfnamefont {V.~R.}\ \bibnamefont
  {Manfrinato}}, \bibinfo {author} {\bibfnamefont {A.}~\bibnamefont {Stein}},
  \bibinfo {author} {\bibfnamefont {L.}~\bibnamefont {Zhang}}, \bibinfo
  {author} {\bibfnamefont {C.-Y.}\ \bibnamefont {Nam}}, \bibinfo {author}
  {\bibfnamefont {K.~G.}\ \bibnamefont {Yager}}, \bibinfo {author}
  {\bibfnamefont {E.~A.}\ \bibnamefont {Stach}}, \ and\ \bibinfo {author}
  {\bibfnamefont {C.~T.}\ \bibnamefont {Black}},\ }\href@noop {} {\bibfield
  {journal} {\bibinfo  {journal} {Nano Letters}\ }\textbf {\bibinfo {volume}
  {17}},\ \bibinfo {pages} {4562} (\bibinfo {year} {2017})}\BibitemShut
  {NoStop}%
\end{thebibliography}%

\end{document}